\documentclass[10pt,aps,prd,twocolumn,twoside,superscriptaddress,floatfix, nofootinbib,amssymb]{revtex4-1}
\usepackage[T1]{fontenc}

\pdfoutput=1
\usepackage{ifthen,amsmath,amssymb, bm}
\usepackage{graphicx}
\usepackage[dvipsnames]{xcolor}
\usepackage{aas_macros}
\usepackage{caption}
\usepackage{subcaption}
\usepackage[none]{hyphenat}
\usepackage{footmisc}
\usepackage[export]{adjustbox}
\usepackage{bookmark}
\usepackage{hyperref}

\usepackage[normalem]{ulem}

\newcommand{\beq} {\begin{equation}}
\newcommand{\eeq} {\end{equation}}
\newcommand{\bal} {\begin{aligned}}
\newcommand{\eal} {\end{aligned}}
\begin{document}

\title{Backlighting extended gas halos around luminous red galaxies: \\
kinematic Sunyaev-Zel'dovich effect from DESI Y1 x ACT}

\author{Bernardita Ried Guachalla}
\email{bried@stanford.edu}
\affiliation{Department of Physics, Stanford University, Stanford, CA, USA 94305-4085}
\affiliation{Kavli Institute for Particle Astrophysics and Cosmology, 382 Via Pueblo Mall Stanford, CA 94305-4060, USA}
\affiliation{SLAC National Accelerator Laboratory 2575 Sand Hill Road Menlo Park, California 94025, USA}
\author{Emmanuel Schaan}
\affiliation{Kavli Institute for Particle Astrophysics and Cosmology,
382 Via Pueblo Mall Stanford, CA 94305-4060, USA}
\affiliation{SLAC National Accelerator Laboratory 2575 Sand Hill Road Menlo Park, California 94025, USA}
\author{Boryana Hadzhiyska}
\affiliation{Physics Division, Lawrence Berkeley National Laboratory, Berkeley, CA 94720, USA}
\affiliation{Berkeley Center for Cosmological Physics, Department of Physics, University of California, Berkeley, CA 94720, USA}
\affiliation{Miller Institute for Basic Research in Science, University of California, Berkeley, CA, 94720, USA}
\author{Simone Ferraro}
\affiliation{Physics Division, Lawrence Berkeley National Laboratory, Berkeley, CA 94720, USA}
\affiliation{Berkeley Center for Cosmological Physics, Department of Physics, University of California, Berkeley, CA 94720, USA}

\author{Jessica N. Aguilar}
\affiliation{Lawrence Berkeley National Laboratory, 1 Cyclotron Road, Berkeley, CA 94720, USA}

\author{Steven Ahlen}
\affiliation{Boston University, 590 Commonwealth Avenue, Boston, MA 02215, USA}
\author{Nicholas Battaglia}
\affiliation{Department of Astronomy, Cornell University, Ithaca, NY 14853, USA}
\affiliation{Université Paris Cité, CNRS, Astroparticule et Cosmologie, F-75013 Paris, France}
\author{Davide Bianchi}
\affiliation{Dipartimento di Fisica ``Aldo Pontremoli'', Universit\`a degli Studi di Milano, Via Celoria 16, I-20133 Milano, Italy}
\author{Richard Bond}
\affiliation{CITA, University of Toronto, Toronto ON M5S 3H8, Canada}
\author{David Brooks}
\affiliation{Department of Physics \& Astronomy, University College London, Gower Street, London, WC1E 6BT, UK}
\author{Todd~Claybaugh}
\affiliation{Lawrence Berkeley National Laboratory, 1 Cyclotron Road, Berkeley, CA 94720, USA}
\author{William~R.~Coulton}
\affiliation{Kavli Institute for Cosmology Cambridge, Madingley Road, Cambridge CB3 0HA, UK}
\affiliation{DAMTP, Centre for Mathematical Sciences, University of Cambridge, Wilberforce Road, Cambridge CB3 OWA, UK}
\author{Axel de la Macorra}
\affiliation{Instituto de F\'{\i}sica, Universidad Nacional Aut\'{o}noma de M\'{e}xico,  Circuito de la Investigaci\'{o}n Cient\'{\i}fica, Ciudad Universitaria, Cd. de M\'{e}xico  C.~P.~04510,  M\'{e}xico}
\author{Mark J. Devlin}
\affiliation{Department of Physics and Astronomy, University of Pennsylvania, 209 South 33rd Street, Philadelphia, PA 19104, USA}
\author{Arjun Dey}
\affiliation{NSF NOIRLab, 950 N. Cherry Ave., Tucson, AZ 85719, USA}
\author{Peter~Doel}
\affiliation{Department of Physics \& Astronomy, University College London, Gower Street, London, WC1E 6BT, UK}
\author{Jo~Dunkley}
\affiliation{Joseph Henry Laboratories of Physics, Jadwin Hall, Princeton University, Princeton, NJ, USA 08544}
\affiliation{Department of Astrophysical Sciences, Peyton Hall, Princeton University, Princeton, NJ, USA 08544}
\author{Kevin~Fanning}
\affiliation{Kavli Institute for Particle Astrophysics and Cosmology, 382 Via Pueblo Mall Stanford, CA 94305-4060, USA}
\affiliation{SLAC National Accelerator Laboratory 2575 Sand Hill Road Menlo Park, California 94025, USA}
\author{Jaime~Forero-Romero}
\affiliation{Departamento de F\'isica, Universidad de los Andes, Carrera 1 N\'umero 18A-10, Edificio Ip, CP 111711, Bogot\'a, Colombia}
\affiliation{Observatorio Astron\'omico, Universidad de los Andes, Carrera 1 N\'umero 18A-10, Edificio H, CP 111711 Bogot\'a, Colombia}
\author{Enrique Gazta\~naga}
\affiliation{Institut d'Estudis Espacials de Catalunya (IEEC), c/ Esteve Terradas 1, Edifici RDIT, Campus PMT-UPC, 08860 Castelldefels, Spain}
\affiliation{Institute of Cosmology and Gravitation, University of Portsmouth, Dennis Sciama Building, Portsmouth, PO1 3FX, UK}
\affiliation{Institute of Space Sciences, ICE-CSIC, Campus UAB, Carrer de Can Magrans s/n, 08913 Bellaterra, Barcelona, Spain}
\author{Satya Gontcho A Gontcho}
\affiliation{Lawrence Berkeley National Laboratory, 1 Cyclotron Road, Berkeley, CA 94720, USA}
\author{Gaston Gutierrez}
\affiliation{Fermi National Accelerator Laboratory, PO Box 500, Batavia, IL 60510, USA}
\author{Julien Guy}
\affiliation{Lawrence Berkeley National Laboratory, 1 Cyclotron Road, Berkeley, CA 94720, USA}
\author{J.~Colin Hill}
\affiliation{Department of Physics, Columbia University, New York, NY, USA 10027}
\author{Klaus Honscheid}
\affiliation{Center for Cosmology and AstroParticle Physics, The Ohio State University, 191 West Woodruff Avenue, Columbus, OH 43210, USA}
\affiliation{Department of Physics, The Ohio State University, 191 West Woodruff Avenue, Columbus, OH 43210, USA}
\affiliation{The Ohio State University, Columbus, 43210 OH, USA}
\author{Stephanie~Juneau}
\affiliation{NSF NOIRLab, 950 N. Cherry Ave., Tucson, AZ 85719, USA}
\author{Theodore~Kisner}
\affiliation{Lawrence Berkeley National Laboratory, 1 Cyclotron Road, Berkeley, CA 94720, USA}
\author{Anthony~Kremin}
\affiliation{Lawrence Berkeley National Laboratory, 1 Cyclotron Road, Berkeley, CA 94720, USA}
\author{Andrew Lambert}
\affiliation{Lawrence Berkeley National Laboratory, 1 Cyclotron Road, Berkeley, CA 94720, USA}
\author{Martin Landriau}
\affiliation{Lawrence Berkeley National Laboratory, 1 Cyclotron Road, Berkeley, CA 94720, USA}
\author{Laurent~Le~Guillou}
\affiliation{Sorbonne Universit\'{e}, CNRS/IN2P3, Laboratoire de Physique Nucl\'{e}aire et de Hautes Energies (LPNHE), FR-75005 Paris, France}
\author{Niall MacCrann}
\affiliation{Kavli Institute for Cosmology Cambridge, Madingley Road, Cambridge CB3 0HA, UK}
\affiliation{DAMTP, Centre for Mathematical Sciences, University of Cambridge, Wilberforce Road, Cambridge CB3 OWA, UK}
\author{Marc Manera}
\affiliation{Departament de F\'{i}sica, Serra H\'{u}nter, Universitat Aut\`{o}noma de Barcelona, 08193 Bellaterra (Barcelona), Spain}
\affiliation{Institut de F\'{i}sica d'Altes Energies (IFAE), The Barcelona Institute of Science and Technology, Edifici Cn, Campus UAB, 08193, Bellaterra (Barcelona), Spain}
\author{Aaron Meisner}
\affiliation{NSF NOIRLab, 950 N. Cherry Ave., Tucson, AZ 85719, USA}
\author{Ramon Miquel}
\affiliation{Institut de F\'{i}sica d'Altes Energies (IFAE), The Barcelona Institute of Science and Technology, Edifici Cn, Campus UAB, 08193, Bellaterra (Barcelona), Spain}
\affiliation{Instituci\'{o} Catalana de Recerca i Estudis Avan\c{c}ats, Passeig de Llu\'{\i}s Companys, 23, 08010 Barcelona, Spain}
\author{Kavilan Moodley}
\affiliation{A. Astrophysics Research Centre, University of KwaZulu-Natal, Westville Campus, Durban 4041, South Africa}
\affiliation{B. School of Mathematics, Statistics \& Computer Science, University of KwaZulu-Natal, Westville Campus, Durban 4041, South Africa}
\author{John Moustakas}
\affiliation{Department of Physics and Astronomy, Siena College, 515 Loudon Road, Loudonville, NY 12211, USA}
\author{Tony Mroczkowski}
\affiliation{European Southern Observatory (ESO), Karl-Schwarzschild-Strasse 2, Garching 85748, Germany}
\author{Adam~D.~Myers}
\affiliation{Department of Physics \& Astronomy, University  of Wyoming, 1000 E. University, Dept.~3905, Laramie, WY 82071, USA}
\author{Michael D. Niemack}
\affiliation{Department of Astronomy, Cornell University, Ithaca, NY 14853, USA}
\affiliation{Department of Physics, Cornell University, Ithaca, NY 14853, USA}
\author{Gustavo Niz}
\affiliation{Departamento de F\'{\i}sica, DCI-Campus Le\'{o}n, Universidad de Guanajuato, Loma del Bosque 103, Le\'{o}n, Guanajuato C.~P.~37150, M\'{e}xico.}
\affiliation{Instituto Avanzado de Cosmolog\'{\i}a A.~C., San Marcos 11 - Atenas 202. Magdalena Contreras. Ciudad de M\'{e}xico C.~P.~10720, M\'{e}xico}
\author{Nathalie~Palanque-Delabrouille}
\affiliation{Lawrence Berkeley National Laboratory, 1 Cyclotron Road, Berkeley, CA 94720, USA}
\affiliation{IRFU, CEA, Universit\'{e} Paris-Saclay, F-91191 Gif-sur-Yvette, France}
\author{Will~Percival}
\affiliation{Department of Physics and Astronomy, University of Waterloo, 200 University Ave W, Waterloo, ON N2L 3G1, Canada}
\affiliation{Perimeter Institute for Theoretical Physics, 31 Caroline Street North, Waterloo, ON N2L 2Y5, Canada}
\affiliation{Waterloo Centre for Astrophysics, University of Waterloo, 200 University Avenue W, Waterloo, ON N2L 3G1, Canada}
\author{Ignasi P\'erez-R\`afols}
\affiliation{Departament de F\'isica, EEBE, Universitat Polit\`ecnica de Catalunya, c/Eduard Maristany 10, 08930 Barcelona, Spain}
\author{Claire~Poppett}
\affiliation{Lawrence Berkeley National Laboratory, 1 Cyclotron Road, Berkeley, CA 94720, USA}
\affiliation{Space Sciences Laboratory, University of California, Berkeley, 7 Gauss Way, Berkeley, CA  94720, USA}
\affiliation{University of California, Berkeley, 110 Sproul Hall \#5800 Berkeley, CA 94720, USA}
\author{Francisco~Prada}
\affiliation{Instituto de Astrof\'{i}sica de Andaluc\'{i}a (CSIC), Glorieta de la Astronom\'{i}a, s/n, E-18008 Granada, Spain}
\author{Frank J. Qu}
\affiliation{Kavli Institute for Particle Astrophysics and Cosmology, 382 Via Pueblo Mall Stanford, CA 94305-4060, USA}
\affiliation{SLAC National Accelerator Laboratory 2575 Sand Hill Road Menlo Park, California 94025, USA}
\author{Graziano Rossi}
\affiliation{Department of Physics and Astronomy, Sejong University, 209 Neungdong-ro, Gwangjin-gu, Seoul 05006, Republic of Korea}
\author{Eusebio Sanchez}
\affiliation{CIEMAT, Avenida Complutense 40, E-28040 Madrid, Spain}
\author{David Schlegel}
\affiliation{Lawrence Berkeley National Laboratory, 1 Cyclotron Road, Berkeley, CA 94720, USA}
\author{Michael~Schubnell}
\affiliation{Department of Physics, University of Michigan, 450 Church Street, Ann Arbor, MI 48109, USA}
\affiliation{University of Michigan, 500 South State Street, Ann Arbor, MI 48109, USA}
\author{Hee-Jong Seo}
\affiliation{Department of Physics \& Astronomy, Ohio University, 139 University Terrace, Athens, OH 45701, USA}
\author{Crist\'obal Sif\'on}
\affiliation{Instituto de F\'isica, Pontificia Universidad Cat\'olica de Valpara\'iso, Casilla 4059, Valpara\'iso, Chile}
\author{David N. Spergel}
\affiliation{Flatiron Institute, 162 Fifth Avenue, NY 10111 USA}
\author{David Sprayberry}
\affiliation{NSF NOIRLab, 950 N. Cherry Ave., Tucson, AZ 85719, USA}
\author{Gregory Tarl\'{e}}
\affiliation{University of Michigan, 500 South State Street, Ann Arbor, MI 48109, USA}
\author{Mariana~Vargas-Maga\~na}
\affiliation{Instituto de F\'{\i}sica, Universidad Nacional Aut\'{o}noma de M\'{e}xico,  Circuito de la Investigaci\'{o}n Cient\'{\i}fica, Ciudad Universitaria, Cd. de M\'{e}xico  C.~P.~04510,  M\'{e}xico}
\author{Eve M. Vavagiakis}
\affiliation{Department of Physics, Cornell University, Ithaca, NY 14853, USA}
\affiliation{Department of Physics, Duke University, Durham, NC 27710, USA}
\author{Benjamin A. Weaver}
\affiliation{NSF NOIRLab, 950 N. Cherry Ave., Tucson, AZ 85719, USA}
\author{Edward J. Wollack}
\affiliation{NASA Goddard Space Flight Center, 8800 Greenbelt Road, Greenbelt, MD 20771, USA}
\author{Pauline~Zarrouk}
\affiliation{Sorbonne Universit\'{e}, CNRS/IN2P3, Laboratoire de Physique Nucl\'{e}aire et de Hautes Energies (LPNHE), FR-75005 Paris, France}

\begin{abstract}

The gas density profile around galaxies, shaped by feedback and affecting the galaxy lensing signal, is imprinted on the cosmic microwave background (CMB) by the kinematic Sunyaev-Zel'dovich effect (kSZ).
We precisely measure this effect ($S/N\approx 10$) via velocity stacking with more than 800,000 spectroscopically confirmed luminous red galaxies (LRG) from the Dark Energy Spectroscopic Instrument (DESI) Y1 survey, which overlap with the Atacama Cosmology Telescope (ACT) Data Release 6 temperature maps over $\geq$ 4,000 deg$^2$.
We explore the kSZ dependence with various galaxy parameters and find no significant trend with redshift, but clear trends with stellar mass and absolute magnitude in $g$, $r$, and $z$ bands.
Our analysis suggests that the gas extends beyond the dark matter halo (99.5\% confidence, i.e. PTE = 0.005).
We find a tentative preference for hydrodynamical simulation models with stronger feedback that drives gas further out (Illustris $z=0.5$, PTE = 0.37) over weaker-feedback cases (IllustrisTNG $z=0.8$, PTE = 0.045), though with limited statistical significance.
In all cases, a free multiplicative amplitude was fit to the simulated profiles, and further modeling work is required to firm up these conclusions.
We find consistency between kSZ profiles around spectroscopic and photometric LRG, with comparable statistical power, thus increasing our confidence in the photometric analysis.
Additionally, we present the first kSZ measurement around DESI Y1 bright galaxy sample (BGS) and emission-line galaxies (ELG), whose features match qualitative expectations.
Finally, we forecast $S/N \sim 50$ for future stacked kSZ measurements using data from ACT, DESI Y3, and Rubin Observatory.
These measurements will serve as an input for galaxy formation models and baryonic uncertainties in galaxy lensing.
\end{abstract}

\maketitle

\section{Introduction}
\label{sec:intro}

The cosmic microwave background (CMB) radiation shows small temperature fluctuations, known as anisotropies. 
On larger scales, we observe primary anisotropies, which carry the imprint of interactions from the early universe, up to the surface of last scattering.
On smaller scales, secondary anisotropies arise from the subsequent interaction of CMB photons with the large-scale structure (LSS) of the Universe. 
Detecting some of these secondary anisotropies is challenging because they could be either faint or have a small angular size.
Fortunately, galaxy surveys allow us to pinpoint and study multiple secondary anisotropies in CMB maps that would otherwise be challenging to detect.
As a result, deeper and more sensitive CMB experiments, combined with cutting-edge galaxy surveys, are essential for detecting and further characterizing these types of secondary anisotropies.

In the last decade, data from Stage 3 CMB experiments, including the \textit{Planck} Satellite \cite{2003Planck, Planck2020}, the Atacama Cosmology Telescope (ACT) \cite{Fowler_2007, Swetz2011, Thornton_2016, Naess_2020, louis2025atacamacosmologytelescopedr6}, and the South Pole Telescope (SPT) \cite{SPT_2004, 2011SPT, Story_2013}, cross-matched with galaxy surveys, such as the Dark Energy Survey (DES) \cite{2016DES, Abbott_2022, Chang_2023} and the Baryon Oscillation Spectroscopic Survey (BOSS) \cite{Dawson_2012, Alam_2017, 2017Singh}, allowed the measurement of many of these secondary anisotropy effects.
For example, we can observe the imprint on the CMB when photons scatter off free electrons in gas in and around clusters of galaxies, the Sunyaev-Zel'dovich effect (SZ) \cite{Sunyaev1980}.

The SZ effect results when CMB photons Compton scatter off the ionized gas in and around galaxies and clusters.
The SZ effect happens either because the free electrons have thermal motion (tSZ) \cite{Zeldovich_Sunyaev_1969, Sunyaev_Zeldovich_1972}, or because there is a relative velocity between the gas and the CMB rest frame along the line-of-sigh (kSZ) \cite{Ostriker_Vishniac_1986, Vishniac_1987}. 
Both effects have been measured, first for individual clusters \cite{1994Birkinshaw, Huang_2020, Hand:2012ui} and more recently, for large numbers of galaxies \cite{Schaan2021, liu2025measurementsthermalsunyaevzeldovicheffect}.

There are different ways of measuring the kSZ signal from galaxies, including, but not limited to, the pairwise method \cite{Hand:2012ui, 2015Hdz-Mont, Bernardis_2017, Soergel_2016, Sugiyama_2018, Calafut_2021, li2024detection}, projected fields \cite{Hill_2016, Ferraro_2016, Kusiak_2021, Bolliet_2022}, matched filter optimization \cite{Li_2014} and, velocity reconstruction stacking \cite{Schaan:2015uaa, Schaan2021, Tanimura_2021, Tanimura_2022, Mallaby_Kay_2023, hadzhiyska2024evidencelargebaryonicfeedback}.
Combined with the tSZ counterpart, it is possible to model the gas thermodynamics \cite{Battaglia_2017, Amodeo2021}.

Measurements of the kSZ effect play a key role in understanding of astrophysics and cosmology.
On the astrophysical side, the kSZ effect provides a means of characterizing the ``missing'' baryons that extend beyond the interstellar medium \cite{1992Persic, Fukugita_2004, 2012Shull, Bullock2017, Battaglia_2017}.
The total mass in stars accounts for only approximately 10$\%$ of the total baryonic content, with the remaining 90$\%$ believed to be distributed in the form of gas. 
This gas is distributed as follows: 15-20$\%$ exists as warm and hot gas within galaxies and clusters, about 25-30$\%$ is in the form of cold intergalactic gas, and the remaining 50-60$\%$ is found as the warm-hot intergalactic medium (WHIM) \cite{Cen_2006}.
Given that the kSZ signal scales linearly with the local free electron density and does not depend on the temperature or clumpiness of the environment, it serves as an excellent probe of the total baryon density. 
Other efforts involve the use of alternative probes, such as fast radio bursts \cite{reischke2024calibratingbaryonicfeedbackweak}, which are limited by the number of observed events, and X-rays \cite{Eckert2016, Akino2022}, where the luminosity does not scale linearly with the host halo mass.

On the cosmology side, the kSZ effect provides a way to study the growth of structure and the total matter density field. 
Understanding the specific distribution of baryons around galaxies is crucial for constraining models of galaxy formation and the effect of feedback \cite{2010Mo, Katz_2018}.
Furthermore, baryonic feedback from hydrodynamical simulations, which have been calibrated over the past decade using X-ray and tSZ observations \cite{Schaye2023, Kugel2023, Henden2018, Henden2020, LeBrun2015, McCarthy2016}, has also started to be compared with kSZ results \cite{Moser2021, Moser2022, https://doi.org/10.48550/arxiv.2307.10919, mccarthy2024flamingocombiningkineticsz, bigwood2024weak, bigwood2025caselargescaleagnfeedback}. 
Moreover, the characterization of baryons at smaller cosmological scales using the kSZ effect has provided valuable insights for weak-lensing analyses \cite{Schneider2022, bigwood2024weak}, which, until recently, had not incorporated information from these scales.
This approach also presents a potential solution to the so-called ``$S_8$ tension'' \cite{Leauthaud2017, Preston2023, Amon2022, hadzhiyska2024evidencelargebaryonicfeedback, salcido2024implicationsfeedbacksolutionss8}.

In the coming decade, we expect to improve the SZ signal-to-noise by an order of magnitude \cite{battaglia2019probing}.
This will come from higher-sensitivity CMB maps by Simons Observatory \cite{Ade2019}, Cerro Chajnantor Atacama Telescope (CCAT) \cite{CCAT_Prime_Collaboration_2022} and CMB-S4 \cite{Abazajian2016} experiments, combined with the large-galaxy surveys by the Dark Energy Spectroscopic Instrument (DESI) \cite{Aghamousa2016}, the Rubin Observatory Legacy Survey of Space and Time (LSST) \cite{lsstsciencecollaboration2009lsst}, the Euclid space telescope \cite{Amendola_2018}, and the Nancy Grace Roman space telescope \cite{Spergel_2015}. 
Additionally, these surveys will detect enough galaxies in multiple redshift bins to characterize the kSZ effect through different cosmic epochs. 
This has already been studied in \cite{Soergel2016, Chaves_Montero_2020, Kusiak_2021, hadzhiyska2024evidencelargebaryonicfeedback}, however, some of these studies combine varying datasets, while in other cases, photometric data was used, which comes with the cost of systematic errors.

In this work, we present the kSZ signal from the DESI Year 1 data release Luminous Red Galaxies (LRG) \cite{desicollaboration2024desi2024iisample} cross-matched with the Atacama Cosmology Telescope (ACT) data through a stacking measurement.
We stack CMB cutouts at the positions of LRG galaxies and weight them by their reconstructed velocities along the line-of-sight (LOS), following \cite{Ried_Guachalla_2024, Hadzhiyska2023}.
We report the redshift dependency of the kSZ measurement by splitting the DESI catalog into bins of redshift, which detail kSZ signal evolution.
Additionally, we use photometric estimates of the stellar mass and absolute magnitude from the DESI imaging Legacy Survey \cite{Zhou_2023}, to determine the dependence of the kSZ signal on these quantities

This paper is organized as follows: 
In Section \ref{sec:theory}, we present the theoretical background for the kSZ effect.
In Section \ref{sec:data}, we introduce the two datasets used for the measurement.
In Section \ref{sec:methodology}, we explain the methodology of our kSZ measurement, consisting of the velocity reconstruction of galaxies and later, the stacking of the CMB cutouts on the galaxy positions.
In Section \ref{sec:results}, we present our fiducial kSZ profiles and their evolution with redshift, stellar mass, and absolute magnitude, and in Section \ref{sec:discussion_and_conclusion} we present our conclusions.


\section{Theory: Kinematic Sunyaev-Zel'dovich effect}
\label{sec:theory}

The CMB spectrum gets distorted on small scales ($\sim$1-10 arcmin) when photons Thomson scatter from free electrons in the moving circumgalactic medium.
The change in temperature of the photons due to the kSZ effect at the LOS direction ($\hat{\mathbf{n}}$) can be related to the number density of electrons $n_e$ and their peculiar velocity $\mathbf{v}_{\rm e}$ as,
\begin{equation}
    \frac{\delta T_{\rm kSZ}(\hat{\mathbf{n}})}{T_{\rm CMB}}
    = - \int \frac{d\chi}{1+z} \hspace{0.1 cm} n_{\rm e}(\chi \hat{\mathbf{n}}, z) \hspace{0.1 cm} \sigma_{\rm T} \hspace{0.05 cm} e^{-\tau} \hspace{0.1 cm} \left(\frac{\mathbf{v}_{\rm e} \cdot \hat{\mathbf{n}}}{c} \right) ,
    \label{eq:T_kSZ_wr_CMB}
\end{equation}
where $\chi$ is the comoving distance, $z$ is the redshift, $\sigma_T$ is the Thomson cross-section of an electron, $c$ is the speed of light, and $\tau$ is the mean integrated optical depth along the LOS of the observer,
\begin{equation}
    \tau = \int \frac{d\chi}{1+z} \hspace{0.1 cm} n_e(\chi \hat{\mathbf{n}}) \hspace{0.1 cm} \sigma_T  .
    \label{eq:tau}
\end{equation}

The optical depth of our gaseous halos is typically much smaller than 1 ($\tau \ll 1$), so we can assume that $e^{-\tau} \approx 1$.
Using this approximation, we can rewrite Eq.~\ref{eq:T_kSZ_wr_CMB} to express the kSZ temperature effect as being proportional to the optical depth and the bulk velocity of the electrons from the gaseous halos along the LOS:
\begin{equation}
    \frac{\delta T_{\rm kSZ}(\hat{\mathbf{n}})}{T_{\rm CMB}} 
    =
    - \tau \left( \frac{v_{\rm e}^{\rm LOS}}{c} \right),
    \label{eq:kSZ_estimator}
\end{equation}
where $v_{\rm e}^{\rm LOS} = \mathbf{v}_{\rm e} \cdot \hat{\mathbf{n}}$ represents the LOS component of the electron bulk velocity.

We point out that there is no temperature dependence in the intensity of the kSZ effect\footnote{This holds only in the non-relativistic approximation \cite{Nozawa1998}.}, as is the case for X-ray \cite{1998ApJ...495...80B,2002ARA&A..40..539R} or tSZ \cite{1972CoASP...4..173S,Sunyaev1980} observations.
However, an estimate of the peculiar velocities $v_e^{\rm LOS}$ is required, which introduces an additional challenge, as shown in Eq.~\ref{eq:kSZ_estimator}. 
This is particularly difficult because, first, we cannot directly measure this quantity, and second, there are multiple sources of systematics to account for, such as redshift space distortions.
We can still approximately ``reconstruct'' the large-scale peculiar-velocity field by associating the galaxy spatial distribution to the gravitational attraction of the LSS itself, as shown in previous works \cite{Ried_Guachalla_2024, Hadzhiyska2023}.
In Sec. \ref{sec:vel_rec}, we further explain the adopted velocity-reconstruction method in detail.


\section{Datasets}
\label{sec:data}

\begin{figure}
    \centering
    \includegraphics[width=0.48\textwidth]{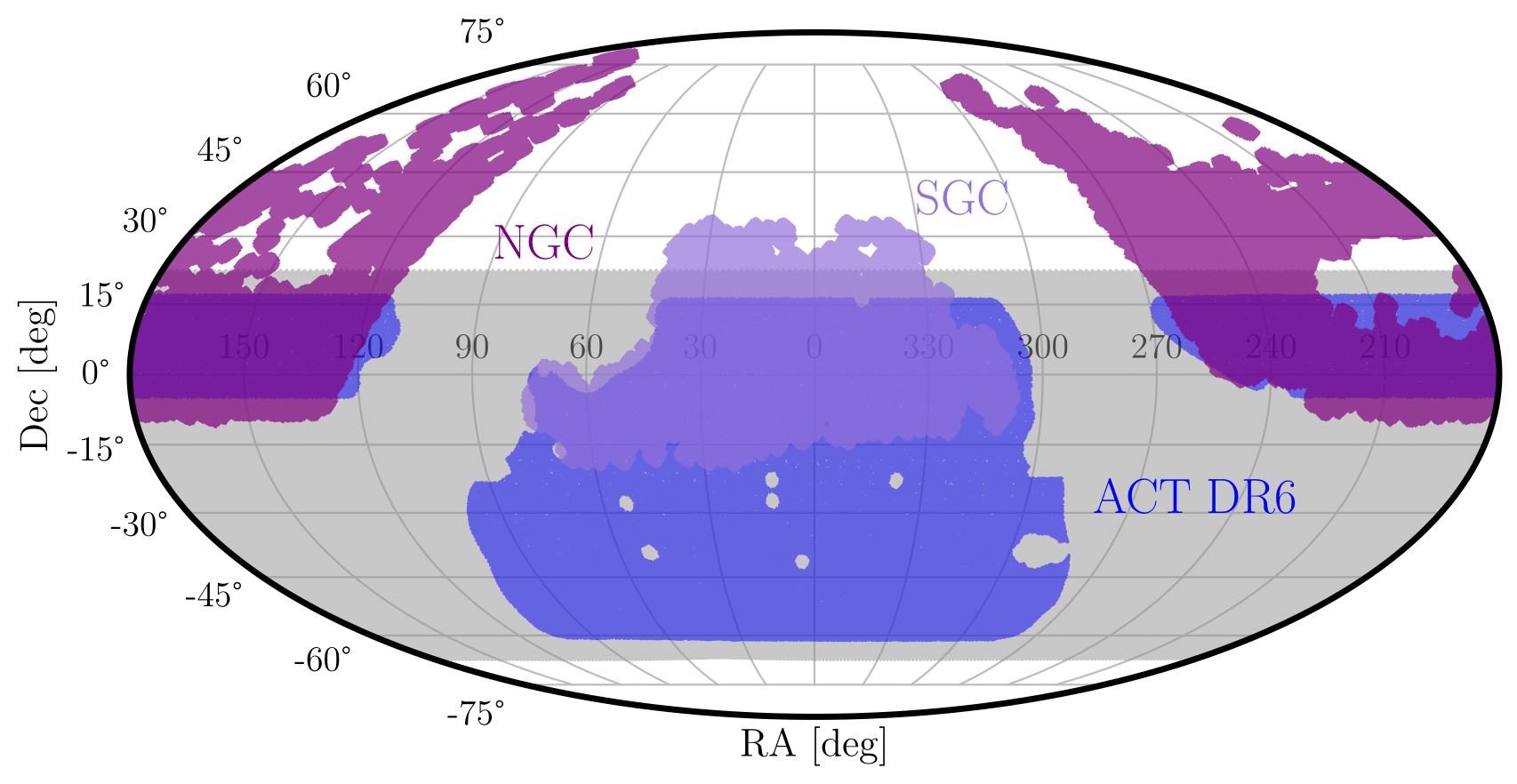}
    \caption{
    ACT DR6 and DESI LRG Y1 footprints in equatorial coordinates.
    In gray, we show the full ACT footprint, while in blue, we show a section of the ACT DR6 survey, where the galactic plane has been masked out, following Planck \cite{Planck2020}. 
    The brightest extended sources and point sources are also masked out, as shown by the small holes visible in the blue area. 
    In purple and violet, we show the North Galactic Cap (NGC) and South Galactic Cap (SGC) samples from DESI Y1, respectively. 
    The SGC overlaps almost fully with the ACT DR6 map, with $\sim78\%$ of the SGC LRG galaxies lying within the ACT footprint, while the NGC overlap is $\sim46\%$.
    The total overlapping region between the two surveys corresponds to $\sim$4,300 deg$^2$.
    }
    \label{fig:footprint}
\end{figure}

For this work, we use two datasets that are displayed in Fig.~\ref{fig:footprint}, the ACT CMB map from data release 6 (DR6) \cite{naess2025atacamacosmologytelescopedr6}, and the partially-overlapping Year 1 spectroscopic luminous-red galaxies (LRG Y1), from DESI \cite{2023Zhou}.


%
%


\subsection{ACT DR6}
\label{sec:act_data}

ACT (2007-2022) was a millimeter wavelength survey covering 18,000 deg$^2$ of the sky \cite{Naess_2020, Thornton_2016}.
It was located on Cerro Toco in the Atacama Desert in northern Chile, and its main goal was to measure the CMB radiation on arcminute scales with high precision.

ACT DR6 consists of temperature and polarization data collected from 2017 to 2021 covering three frequency bands: f090 (77–112 GHz), f150 (124–172 GHz), and f220 (182–277 GHz), where the CMB signal dominates.
We use the night-time observations, using the harmonic-space Internal Linear Combination (hILC) maps (dr6.01 version) produced by combining multiple frequencies of observations from ACT and the Planck satellite \cite{2024Coulton, naess2025atacamacosmologytelescopedr6, louis2025atacamacosmologytelescopedr6, 2003Planck}.
These maps have a pixel of size 0.5 arcmin and are convolved with a Gaussian beam with a FWHM equivalent to 1.6 arcmin.

To avoid contamination on the kSZ measurement arising from foreground clusters and point sources, we apply a mask, which is visible in Fig.~\ref{fig:footprint}.
Similar to \cite{Schaan2021}, we also remove parts of the ACT map that contain temperature outliers exceeding $\pm 5\sigma$ relative to the mean CMB, in order to reduce potential tSZ contamination.


\subsection{DESI Y1 spectroscopic LRG}
\label{sec:desi_data}

DESI is the largest galactic spectroscopic survey currently operating, covering 14,000 deg$^2$ ($\sim 1/3$ of the sky) \cite{DESI2022.KP1.Instr}.
Located at Kitt Peak National Observatory in Arizona, at the 4-meter Mayall telescope, it will obtain the spectra of more than 60 million galaxies and quasars in five years \cite{2022Abareshi, Aghamousa2016, DESI2016b.Instr, DESI2016a.Science, FocalPlane.Silber.2023, Corrector.Miller.2023, FiberSystem.Poppett.2024}.
This is an order of magnitude higher than what previous surveys like BOSS \cite{Eisenstein_2011, Dawson_2012} and eBOSS \cite{2017Blanton, 2020Ahumada} achieved in over a decade of observations.
It will span redshifts of $0.1 < z < 1.6$ for galaxies, and $0.9 < z < 3.5$ and beyond for quasars, covering the evolution of LSS through 10 billion years.

The goal of DESI is to determine the nature of dark energy through the most precise measurement of the expansion history of the universe ever obtained \cite{Snowmass2013.Levi}. 
The final galaxy catalog of DESI will include 8 million LRG galaxies \cite{Spectro.Pipeline.Guy.2023, SurveyOps.Schlafly.2023}, which are particularly advantageous to study for two reasons. 
First, LRG galaxies are highly-biased tracers of LSS, allowing us to measure baryon acoustic oscillations (BAO) \cite{2024DESI_BAO}, and derive cosmological constraints \cite{desicollaboration2024desi}.
Second, their selection is robust due to distinctive spectra \cite{Zhou_2023}.
Additionally, LRG galaxies have a low satellite fraction compared to the other DESI samples ($\sim11\%$, calculated by \cite{yuan2023desi} using the AbacusHOD in the DESI One-Percent Survey \cite{Zhou2020}) and are, therefore, less impacted by observational effects such as off-centering and redshift space distortions (RSD) due to the virial motion of satellites. 
This makes them natural candidates for studying the kSZ effect in the DESI survey.

\begin{figure}
    \centering
    \includegraphics[width=0.48\textwidth]{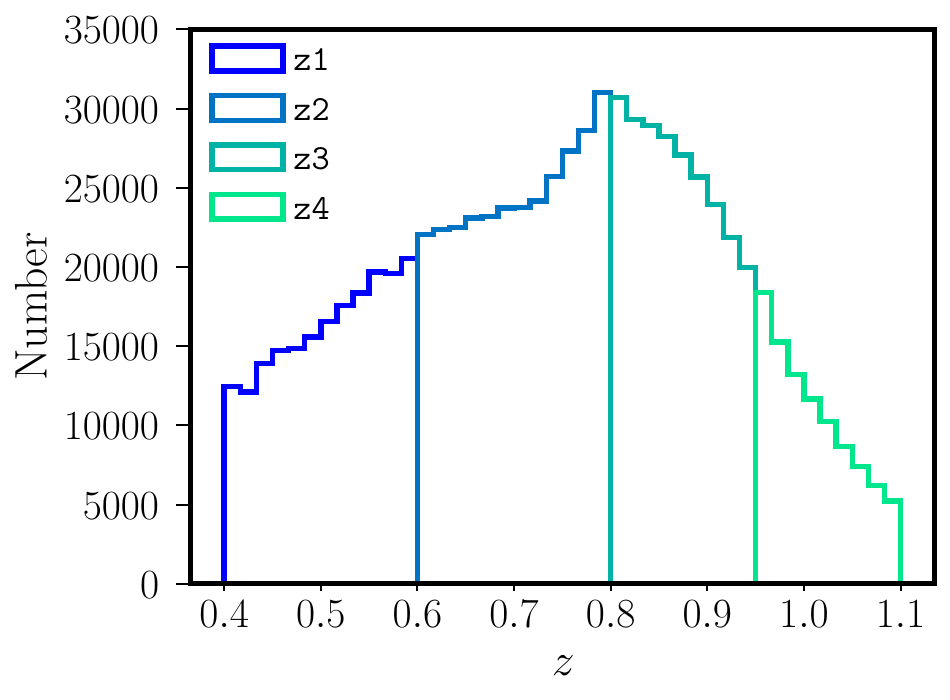}
    \caption{
    Redshift distribution of the DESI LRG Y1 galaxies overlapping the ACT map ($\sim 39\%$ of the total LRG sample).
    The galaxies were divided into four bins given their spectroscopic redshift, shown by the vertical lines.
    Additional summary information, including mean redshift and number of objects per bin, is included in Table~\ref{tab:result_reports}.
    }
    \label{fig:redshift_distribution}
\end{figure}

All DESI target selection is based on the public Legacy Surveys \cite{LS.Overview.Dey.2019} and has been validated during the DESI Survey Validation phase \cite{DESI2023a.KP1.SV, Zhou2020}. 
Following the validation phase, the Early DESI Data Release \cite{DESI2023b.KP1.EDR}, the first measurements of the BAO signals in galaxies \cite{BAO.EDR.Moon.2023} and the Lyman alpha forest \cite{2023JCAP...11..045G} were made.

The first data release (DR1) \cite{DESI2024.I.DR1}, data we will use in this work, includes several key science papers presenting two-point clustering measurements and validation \cite{DESI2024.II.KP3}, BAO measurements from galaxies and quasars \cite{DESI2024.III.KP4}, and from the Lyman alpha forest \cite{DESI2024.IV.KP6}, as well as a full-shape study of galaxies and quasars \cite{DESI2024.V.KP5}. 
Additionally, there are cosmological results from the BAO measurements \cite{DESI2024.VI.KP7A} and the full-shape analysis \cite{DESI2024.VII.KP7B}.

The DESI LRG Y1 galaxies overlapping with ACT are distributed in $\sim$4,300 deg$^2$ and span a redshift range of $0.4 < z < 1.1$, as shown in Fig.~\ref{fig:redshift_distribution}. 
This corresponds to doubling the redshift range with respect to previous studies using BOSS and ACT data \cite{Schaan2021} (see Fig.~\ref{fig:z_dist}).
In this work, we use the North and South Galactic Cap datasets (NGC and SGC, respectively) overlapping with the ACT DR6 footprint, as shown in Fig.~\ref{fig:footprint}.
Additionally, we consider the galaxies that do not have a bright point source (such as stars, quasars, and radio sources) within an angular distance of 6$'$.
This corresponds to a final selection of $\sim 39\%$ of the DESI LRG Y1 (825,283 galaxies).
For these galaxies, we study possible correlations, including a kSZ dependence on redshift, stellar mass, and absolute magnitude.
For this work, we use the estimate of the mean halo mass of the LRG galaxy sample provided by \cite{yuan2023desi}, which is approximately $\sim 10^{13.4} M_{\odot}/h$.

In order to detect the redshift evolution of the kSZ effect, we split the DESI LRG Y1 overlapping with ACT into four spectroscopic redshift bins:
$(0.4, 0.6), (0.6, 0.8), (0.8, 0.95) \ {\rm and} \ (0.95, 1.1)$, containing 195,877, 297,440, 235,620 and 96,346 galaxies, respectively, as shown in Fig.~\ref{fig:redshift_distribution}\footnote{We followed the DESI LRG Y1 spectroscopic bins: 
Fig. 1 from \cite{2024DESI_BAO} shows that the lowest redshifts bins ($0.4 < z < 0.6$ and $0.6 < z < 0.8$) have roughly a constant number density around $3.7 \cdot 10^{-4}$ ($h/$Mpc)$^3$. 
Due to the drastic reduction in number density, and following \cite{yuan2023desi}, the higher-redshift sample ($0.8 < z < 1.1$) is split into two samples (from $0.8 < z < 0.95$ and $0.95 < z < 1.1$).}.

\begin{figure}
    \centering
    \includegraphics[width=0.45\textwidth]{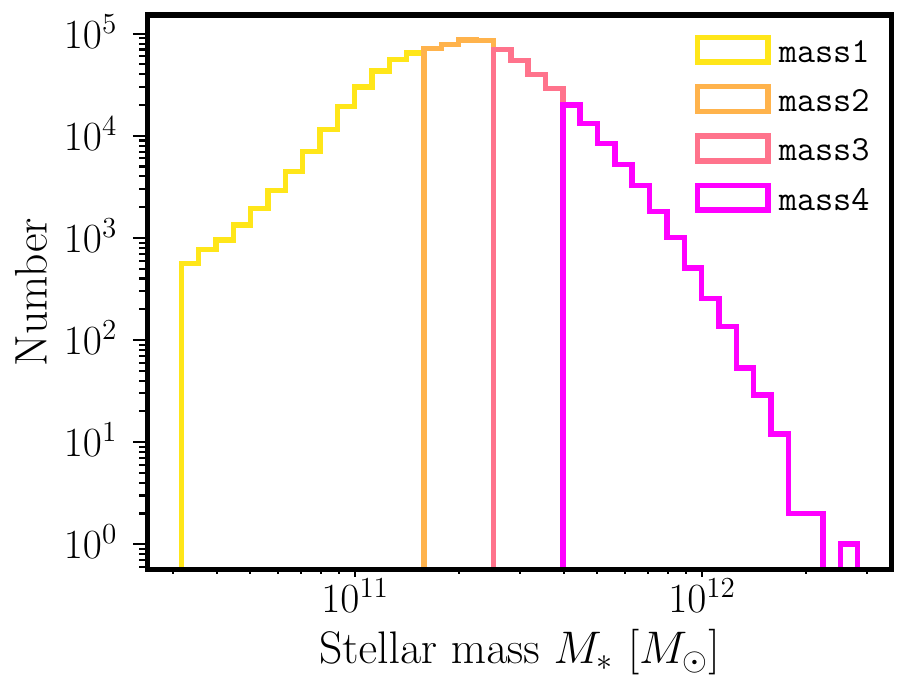}
    \caption{
    The binned stellar mass distribution of the DESI LRG Y1 galaxies overlapping the ACT DR6 map, as estimated by \cite{Zhou_2023}, is provided. 
    Additional summary information for the bins is available in Table~\ref{tab:result_reports}.
    }
    \label{fig:stellar_mass_distribution}
\end{figure}

Additionally, following \cite{hadzhiyska2024evidencelargebaryonicfeedback}, we split the stellar masses into four bins of $\log_{10} (M_{*}/M_{\odot})$: (10.5, 11.2), (11.2, 11.4), (11.4, 11.6), and (11.6, 12.5), as shown in Fig.~\ref{fig:stellar_mass_distribution}.
These bins contain 244,932, 320,914, 194,037, and 53,997 galaxies, respectively\footnote{These bins were determined by optimizing the limits to ensure that the signal-to-noise ratio was sufficient to resolve a dependence.\label{refnote}}.
We use the stellar mass estimates from \cite{Zhou_2023}, which were obtained using a random forest algorithm trained on the DESI Legacy Imaging Survey photometry, with the additional input of stellar masses from \cite{Bundy2015}.

\begin{figure}
    \centering
    \includegraphics[width=0.48\textwidth]{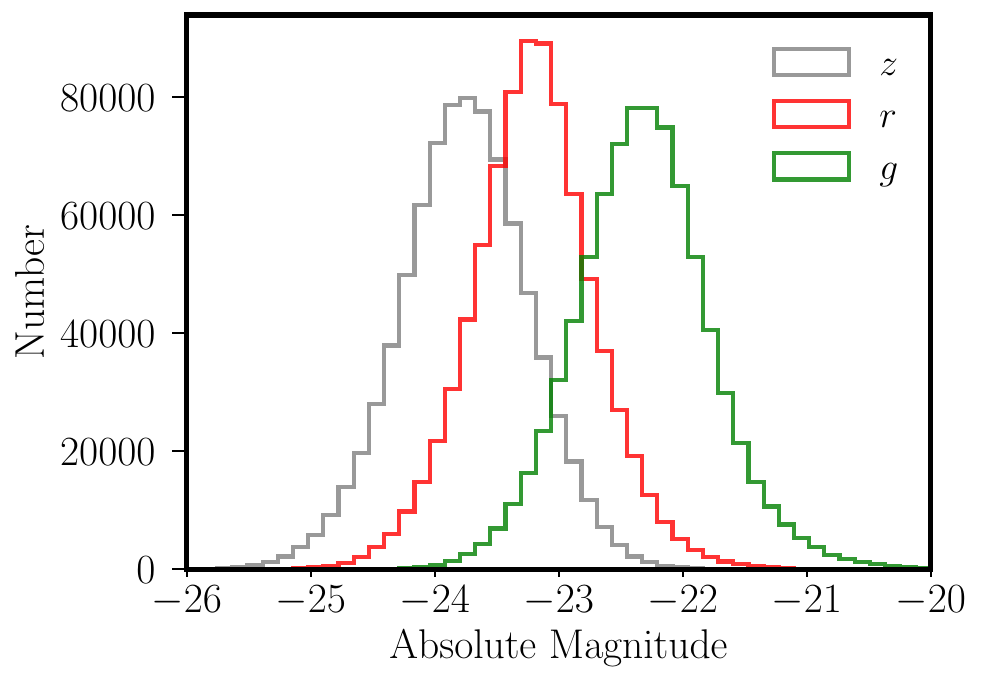}
    \caption{
    Absolute magnitude distribution of the DESI LRG galaxies for the photometric bands $g$ (in green), $r$ (in red), and $z$ (in gray).
    These were obtained by matching the target IDs of the spectroscopic galaxies with the DESI Imaging Legacy Survey \cite{Zhou_2023} and then converting the flux to absolute magnitude per band.
    }
    \label{fig:abs_mag}
\end{figure}

Finally, we analyze the response of the kSZ measure when splitting the sample by their absolute magnitude (after applying a k-correction) for a given band\footref{refnote}:

\begin{itemize}
    \item For the $g$ band, we split into four bins: $M_g \in (-24, -22.5), (-22.5, -22.2), (-22.2, -21.9)$, and $(-21.9, -20)$, containing 299,592, 189,580, 164,457 and 168,848 galaxies, respectively.
    \item For the $r$ band, we split into four bins: $M_g \in (-25, -23.6), (-23.6, -23.2), (-23.2, -22.8)$, and $(-22.8, -21)$,  containing 164,927, 250,337, 250,628 and 159,688 galaxies, respectively.
    \item For the $z$ band, we split into four bins: $M_g \in (-26, -24), (-24, -23.5), (-23.5, -23)$, and $(-23, -22)$, containing 257,941, 315,034, 194,551 and 58,054 galaxies, respectively.
\end{itemize}
For an insight into the distribution of absolute magnitudes, see Fig.~\ref{fig:abs_mag}.

%


\section{Methodology}
\label{sec:methodology}

The methodology for measuring the kSZ effect consists of two parts: 
first, the velocity reconstruction of the gaseous halos, explained in Sec.~\ref{sec:vel_rec}, and second, the stacking of the CMB map at the positions of the galaxies, explained in Sec.~\ref{sec:stacking}.


\subsection{Velocity reconstruction}
\label{sec:vel_rec}

The galaxy number overdensity, $\delta_{\rm g}$, characterizes the clustering of galaxies in the LSS. 
This quantity is related to the matter overdensity, $\delta$, through the linear bias approximation, $\delta_{\rm g} \approx b \delta$, where $b$ is the linear bias factor.
We estimate the 3D peculiar velocity field $\mathbf{v}$ of galaxies by solving the linearized continuity equation:
\begin{equation}
    \nabla \cdot \mathbf{v} = -a H f \frac{\delta_{\rm g}}{b},
    \label{eq:continuity_eq}
\end{equation}
where $a$, $H$ and $f$ are the scale factor, the Hubble parameter and the logarithmic growth rate of structure respectively.
In particular, the logarithmic growth rate can be related to the linear growth factor $D(a)$ by
$f = d \ln D(a) / d \ln a$.
In the presence of RSD, Eq.~\ref{eq:continuity_eq} transforms to \cite{Kaiser1987, Hamilton1998}:
\begin{equation}
    \nabla \cdot \mathbf{v} + \frac{f}{b} \nabla \cdot [(\mathbf{v} \cdot \hat{\mathbf{n}}) \hat{\mathbf{n}}] = -a H f \frac{\delta_{\rm g}}{b},
    \label{eq:continuity_eq_with_RSD}
\end{equation}
where $\mathbf{v} \cdot \hat{\mathbf{n}} = v^{\rm LOS} \approx v^{\rm LOS}_{\rm e}$, is the bulk velocity of the electrons as defined in Eq.~\ref{eq:kSZ_estimator}. 
We note that this is an approximation, since electrons can also have internal motions within halos (e.g., due to recent mergers) that are not captured by the bulk velocity term.
The standard method for solving Eq.~\ref{eq:continuity_eq_with_RSD} is to use numerical techniques in Fourier space. This method can naturally incorporate the Lagrangian displacement field $\pmb{\psi}(\mathbf{x})$, as it is related to the peculiar velocity field,
\begin{equation}
    \mathbf{v}(\mathbf{x}) = aHf \pmb{\psi}(\mathbf{x}).
    \label{eq:vel_distr}
\end{equation}

Determining the displacement field $\pmb{\psi}(\mathbf{x})$ is a fundamental step in clustering analyses.
The BAO signal, a fundamental probe of the early oscillations of the baryon-photon plasma, gets smeared out and mis-centered as non-linear gravitational evolution displaces galaxies on top of the cosmic flow.
To minimize this effect, \cite{Eisenstein_2007, Padmanabhan_2012} proposed to correct for the linear displacement following Eqs.~\ref{eq:continuity_eq_with_RSD} and \ref{eq:vel_distr}.
Nowadays, this correction is widely applied in BAO studies (for BOSS see \cite{VargasMagaa2017, Vargas_Maga_a_2018, Alam_2017}, for DES see \cite{descollaboration2024darkenergysurvey21}, and for DESI Y1 see \cite{2024Paillas, 2024Chen, 2024DESI_BAO}).
The velocity field obtained from Eq.~\ref{eq:vel_distr} has also been used in previous kSZ measurements (e.g., \cite{Schaan2021, Mallaby_Kay_2023, Hadzhiyska2023}).

\begin{figure}
    \centering
    \includegraphics[width=0.48\textwidth]{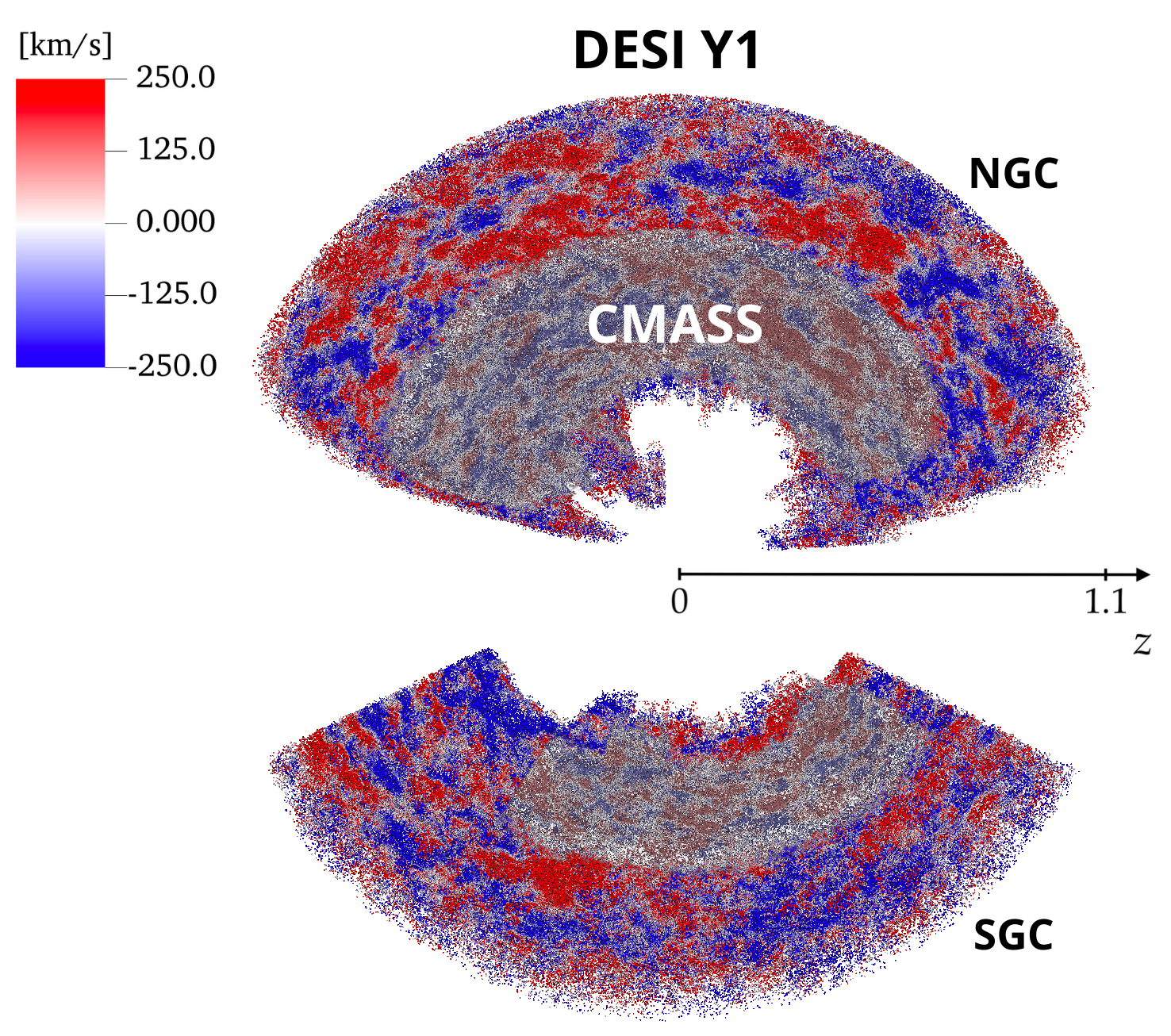}
    \caption{
    Visualization of the reconstructed LOS peculiar velocities of the DESI LRG Y1 using the linear velocity reconstruction method presented in Sec. \ref{sec:vel_rec}.
    On top of the DESI sample, we include the LOS reconstructed velocities of the CMASS (constant mass) sample from the Baryon Oscillation Spectroscopic Survey \cite{Alam2017, Schaan2021, Ahn2014} (DESI's predecessor) in a semi-transparent overlay to show the increment in distance and total number of galaxies.
    Each dot represents a galaxy, and its color will be blue (red) if it moves towards (away from) the observer in the center.
    As in Fig.~\ref{fig:footprint}, NGC and SGC stand for North and South Galactic Cap, respectively, and $z$ stands for redshift.
    }
    \label{fig:vel_rec}
\end{figure}

In this work, we approximate the peculiar velocity field using the BAO reconstruction displacement from the DESI Y1 analysis \cite{2024DESI_BAO}.
Their pipeline includes the use of state-of-the-art mock catalogs mimicking the DESI survey \cite{2024Paillas} and the blinding of the DESI LRG catalog to avoid confirmation bias \cite{andrade2024validating}\footnote{Their reference values for the BAO reconstruction were:
effective redshift $z_{\rm eff}$ = 0.78, smoothing scale $r_s$ = 15 Mpc/$h$, and cosmological parameters $h$ = 0.6736, $\Omega_m$ = 0.31519, and linear bias $b$ = 2.0, from Planck 2018 \cite{Planck2020}.}.
In their pipeline, they implement the symmetric reconstruction (\textsc{RecSym}) reconstruction convention \cite{White2015}, which includes a $1+f$ factor in the displacement along the LOS of galaxies.
This additional step should not be incorporated in our case, as we aim to access the original positions of the galaxies.
We divide by the $1+f$ factor along the LOS displacement of the LRG galaxies and calculate their velocities using Eq.~\ref{eq:vel_distr}.

In Fig.~\ref{fig:vel_rec}, we show the reconstructed radial velocities of the full DESI LRG Y1 sample.
The blue and red colors represent galaxies moving towards and away from an observer at $z=0$, respectively.
Similar to Fig. 1 from \cite{Schaan2021}, we find large-scale velocity structures extended transversely to the LOS.
In Fig.~\ref{fig:vrec_vcent_vsat_1D_histogram}, we display the reconstructed radial-velocity distribution derived from the same data as in Fig.~\ref{fig:vel_rec}.
Overall, the resulting distribution, in gray, is close to a Gaussian with tails, as found in simulations \cite{Ried_Guachalla_2024}.
Nevertheless, the distribution is slightly negatively skewed (note that the plot is on a logarithmic scale, which magnifies the skewness visually).
When examining the sample in detail, we find a total of 49 galaxies with $v^{\rm LOS} < -5 \sigma^{\rm rec}_v = -1,150$ km/s and 25 galaxies with $v^{\rm LOS} > 5 \sigma^{\rm rec}_v = 1,150$ km/s. 
In App.~\ref{sec:skewness} we study the origin of this skewness and find that it is produced by the galaxies at the edges of the redshift distribution ($z$ near 0.4 and 1.1).
In other words, it is generated by the periodic boundary conditions of the reconstruction method (\cite{2024DESI_BAO} uses the \texttt{MultiGrid} \cite{White2015} technique via the package \texttt{pyrecon}\footnote{\url{https://github.com/cosmodesi/pyrecon}}).
In black we show the reconstructed velocities along the LOS, when excluding the edges, resulting in a symmetric distribution.
We note this skewness does not impact our kSZ measurement, as the number of galaxies affected is much lower than their total number.


\subsection{Stacking weighted by the reconstructed peculiar velocities}
\label{sec:stacking}

The kSZ effect is small compared to the CMB primary anisotropies and, therefore, stacking over a large number of galaxies is required for a significant detection on the CMB maps.
The typical order of magnitude of the kSZ effect at the scales where it peaks is $T_{\rm kSZ} \sim 10^{-1} \mu$K \cite{Hu:2001bc}.
For comparison, the CMB primary fluctuations are $100 \mu $K \cite{Fixsen_1996}.
Following \cite{Schaan2021}, we stack galaxies using the compensated-aperture photometry (CAP) filtering technique to study the radial distribution of the kSZ effect around galaxies, previously implemented in the \textsc{ThumbStack}\footnote{\url{https://github.com/EmmanuelSchaan/ThumbStack}} package \citep{2021Schaanthumbstack}.

\begin{figure}
    \centering
    \includegraphics[width=0.48\textwidth]{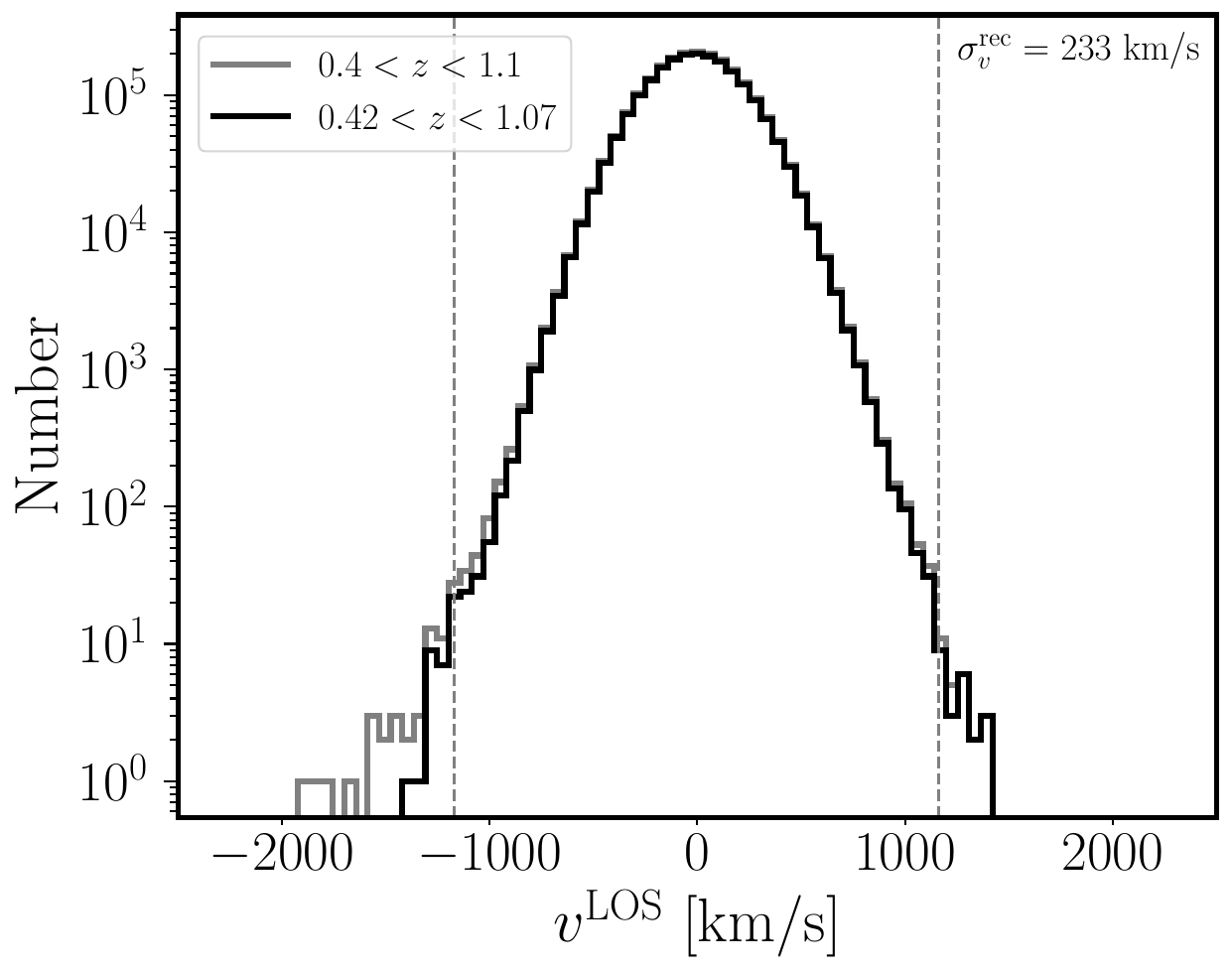}
    \caption{
    In gray, histogram of the reconstructed LOS velocities of the DESI LRG Y1.
    The velocities were obtained from the displacement field $\pmb{\psi}(\mathbf{x})$ of the DESI Y1 BAO measurement \cite{2024Paillas}.
    The quantity $\sigma_v^{\rm rec}$ corresponds to the root-mean-square (RMS) velocity along the LOS and the dashed vertical lines corresponds to $\pm 5 \sigma_v^{\rm rec}$.
    In black, we show the same distribution, but only for galaxies within $0.42<z<1.07$.
    By considering this sub-sample, the slight skewness on the left is removed, indicating the presence of a small survey volume effect, as discussed in App.~\ref{sec:skewness}. 
    We find that this effect does not impact our results.
    }
    \label{fig:vrec_vcent_vsat_1D_histogram}
\end{figure}

First, for varying aperture radius, we measure the inner integrated temperature fluctuation $\mathcal{T}(R)$ as
\begin{equation}
  \mathcal{T}(R) =
\int_0^{R} d^2\theta \hspace{0.05 cm} \delta T(\theta)
- \int_{R}^{\sqrt{2} R} d^2\theta \hspace{0.05 cm} \delta T(\theta).
    \label{eq:CAP}
\end{equation}

Eq.~\ref{eq:CAP} selects the temperature fluctuation of an inner circle of radius $R$, and then subtracts an outer ring of the same area (see Fig. 6 from \cite{Schaan2021} for a visual representation).
The resulting CAP filter, measured at the positions of the galaxies and properly weighted by the velocities along the LOS, would resemble a cumulative gas density profile. As the disk radius $R$ increases, up to the point where the entire gas profile is enclosed within the disk, the output approximates the integrated gas profile.

Second, at the position of each galaxy, we stack the inverse-variance weighted CAP filters for a given radius $R$, and obtain a minimum-variance unbiased linear estimate of the kSZ effect,
\begin{equation}
    \hat{T}_{\rm kSZ} (R)
    = - \frac{1}{r} 
    \frac{\sigma_v^{\rm rec}}{c}
    \frac{\sum_i \mathcal{T}_i(R) (v^{\rm rec}_i/c)}{\sum_i (v^{\rm rec}_i/c)^2},
    \label{eq:kSZ_stacking}
\end{equation}
where $r$ corresponds to the correlation coefficient between the true velocities and estimated velocities, and $\sigma_v^{\rm rec}$ is the root-mean-square of the reconstructed velocities shown in Fig.~\ref{fig:vrec_vcent_vsat_1D_histogram}.
Without the velocity weighting, the kSZ signal would cancel out, since $\mathcal{T}_i \propto v_{\rm e}^{\rm LOS}$, and this velocity is equally likely to point towards or away from us.
Furthermore, the resulting stacked signal from ACT would instead largely trace the tSZ, and could include contributions from radio and CIB contamination that would not be canceled through the above procedure used to recover the kSZ signal.

From our previous work on simulations \cite{Hadzhiyska2023}, we find that $r = 0.65$ for a spectroscopic survey like DESI.
For the purposes of this paper, we neglect the estimated uncertainty in $r$, as it corresponds to an effect of approximately 2$\%$ (see row 20 of Table 1 in \cite{Hadzhiyska2023}).
In Eq.~\ref{eq:kSZ_stacking}, we also assumed that the CMB map noise is uniform throughout the sky, which differs from what was adopted in \cite{Schaan2021}.

To statistically quantify our measurement, we approximate the likelihood of the data as Gaussian, allowing us to evaluate the significance of the kSZ using a $\chi^2$ statistic for the data vector, $d$, the covariance matrix, C, and model, $m$:
\begin{equation}
    \chi^2_{\rm model} = (d - m)^\top {\rm C}^{-1} \hspace{0.1 cm} (d - m).
    \label{eq:chi_squared}
\end{equation}
In particular, for the null hypothesis of no kSZ signal ($m=0$),
Eq.~\ref{eq:chi_squared} becomes 
$\chi^2_{\rm null} = d^\top {\rm C}^{-1}  \hspace{0.1 cm} d$.
An approximation of the covariance matrix, C, of this measurement is obtained from bootstrap resampling our catalog 10,000 times, which is automatically calculated when using the \textsc{ThumbStack} package.
See App.~\ref{sec:cov_matrix} for more details.

Alternatively, if $m$ contains a template profile with a free amplitude parameter, this can be fitted by minimizing $\chi^2_{\rm model}$, yielding $\chi^2_{\rm best-fit}$ (the maximum-likelihood estimation).
The signal-to-noise ratio for this case is given by:
\begin{equation}
    S/N = \sqrt{\chi^2_{\rm null} - \chi^2_{\rm best-fit}}.
    \label{eq:SNR_null}
\end{equation}
%


\section{Results: measured \texorpdfstring{\lowercase{k}}{k}SZ profiles}
\label{sec:results}

In this section, we describe our resulting kSZ CAP profiles from the DESI LRG Y1 sample that overlaps with ACT DR6.
We start with the fiducial result (using the full available sample) in Sec. \ref{sec:complete} and compare it with the profiles obtained from simulations.
We then explore the dependence on redshift, stellar mass, and absolute magnitude in the following sections.
Lastly, we compare our results when using photometric galaxies instead.

To assess the robustness of our measurements, we perform consistency checks and null tests, including randomly shifting the velocities of galaxies and using a variety of derived maps from ACT.
We show our method passes those tests in App.~\ref{sec:syst_checks}.

\subsection{Fiducial case: full DESI LRG Y1 sample}
\label{sec:complete}

\begin{figure}
    \centering
    \includegraphics[width=0.48\textwidth]{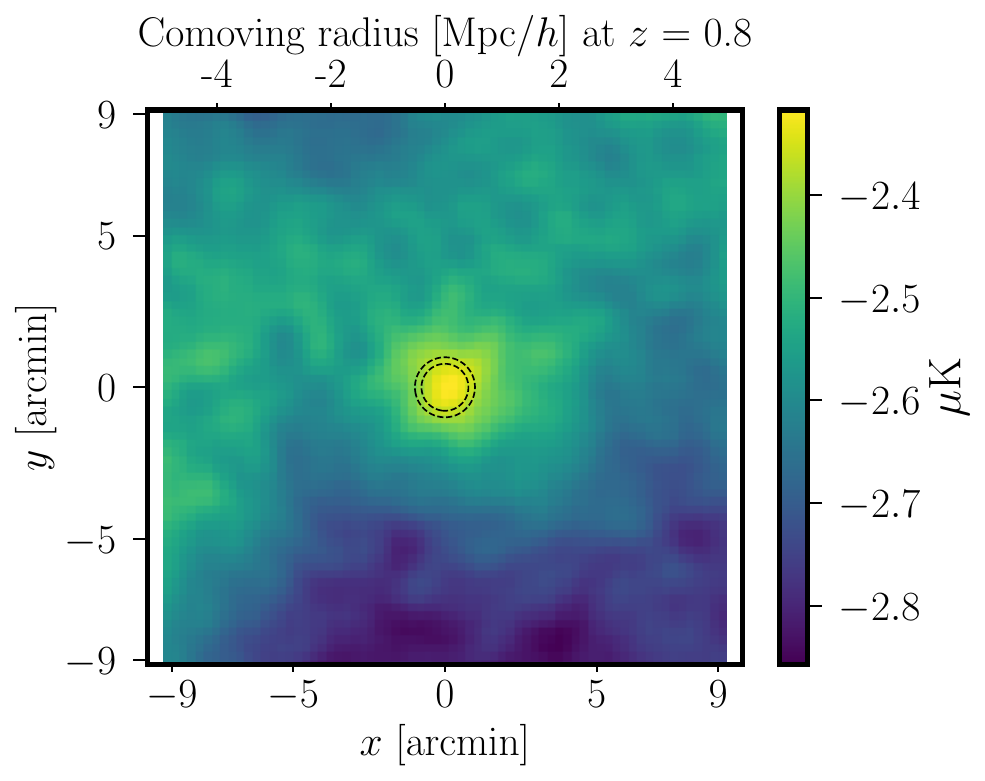}
    \caption{
    For visualization purposes: stacked cutout of the CMB map centered at the galaxy positions and weighted by the reconstructed velocities (Eq.~\ref{eq:kSZ_stacking}), without applying the CAP filter.
    The traced gas density extends radially much more than the halo virial radius (0.56 Mpc/$h$ radius of the inner black circle, corresponding to $\sim$1.2 arcmin) and the beam FWHM (1.6 arcmin, diameter of the outer black circle).}
    \label{fig:stack_map}
\end{figure}

In Fig.~\ref{fig:stack_map}, we show the stacked 2D map made by applying Eq.~\ref{eq:kSZ_stacking} to the cutout around each DESI galaxy, which gives information on the gas density.
Similar to \cite{Schaan2021}, we find that the gas is primarily distributed within a radius of approximately 3.5 Mpc.
For an effective median redshift of  ${\rm Med}(z) \approx 0.8$, this corresponds to an angular size of less than $\sim$6 arcminutes.
The extent of the signal is evident: the inner and outer black circles represent the virial radius (0.56 Mpc/$h$, calculated using $\log(M_h/M_{\odot}) \sim 13.24$ \cite{yuan2023desi}) and the FWHM beam (1.6 arcmin), respectively, showing that the kSZ signal detected spans much larger radii.

\begin{figure}
    \centering
    \includegraphics[width=0.48\textwidth]{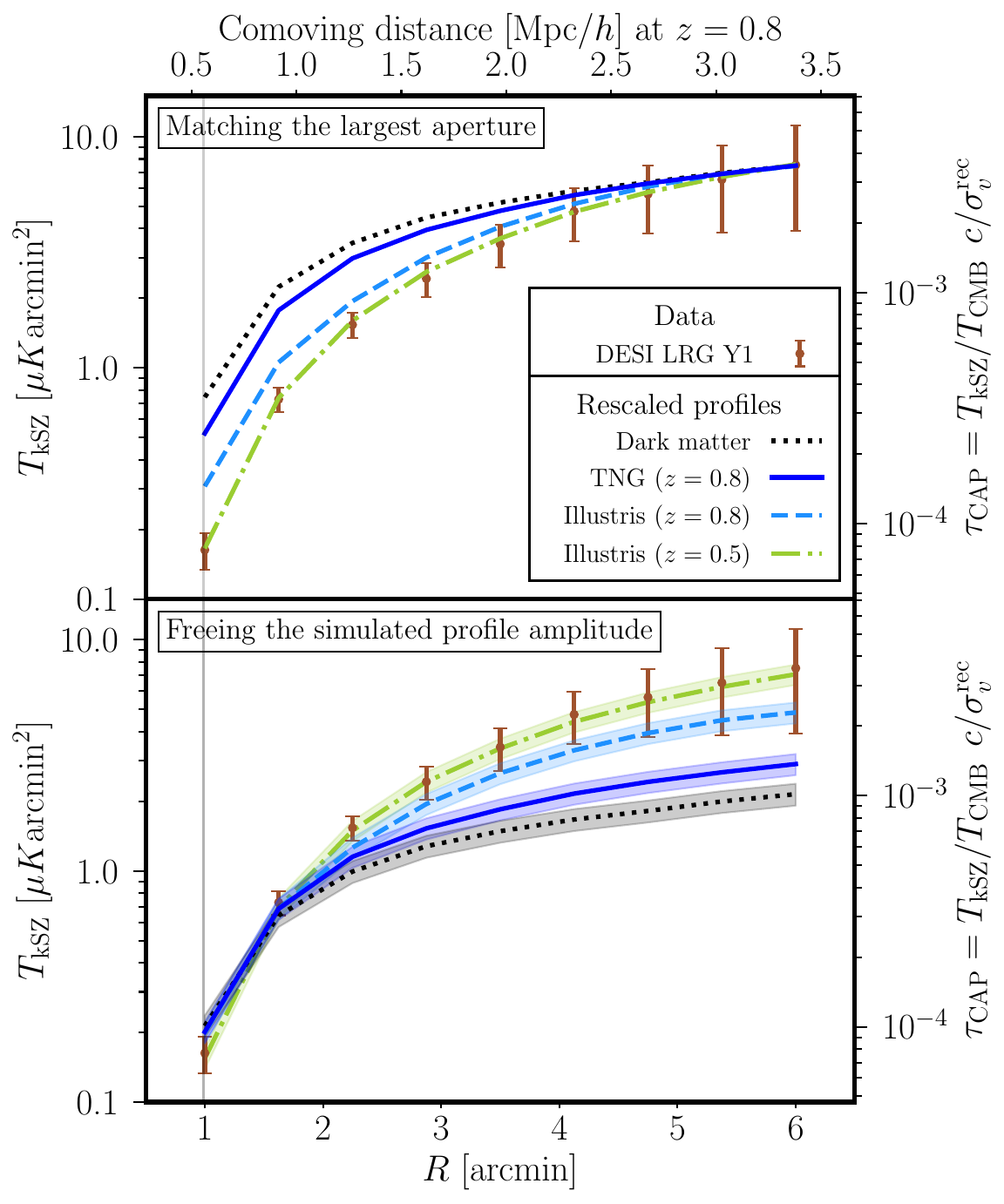}
    \caption{
     The measured stacked kSZ in $\mu K$ arcmin$^2$ for varying CAP filters with radius $R$ from Eq.~\ref{eq:kSZ_stacking} in brown.
     We include the corresponding comoving distances at $z=0.8$ on the top horizontal axis.
     Several simulation CAP profiles are included, each rescaled in amplitude to match the measured signal at the largest aperture where all the baryons should be encompassed (upper panel), or with the amplitude left free to facilitate a comparison of their shapes in the presence of mass mismatch between data and simulation (lower panel). 
     For the IllustrisTNG case (solid blue, labeled TNG in the figure), the profile shape more closely follows that of dark matter than the observed data. 
     For Illustris (light blue dashed) at $z = 0.8$, the profiles tend to align more closely with the observations.
     In contrast, when comparing the Illustris profile at $z=0.5$ (light green dashed), taken from \cite{hadzhiyska2024evidencelargebaryonicfeedback}, we find a better match with the shape of the kSZ profile.
     The bands on the lower panel propagate the uncertainty on the profile amplitude from Eq.~\ref{eq:error_matched_filter}.
     The vertical gray line shows the virial radius added in quadrature with the beam standard deviation (FWHM/$\sqrt{8 \ln(2)}$) and the secondary axis on the right translates to the integrated optical depth of Thomson scattering.
    }
    \label{fig:kSZ_signal}
\end{figure}

In Fig.~\ref{fig:kSZ_signal}, we present the mean stacked kSZ CAP profile (brown error bars) in $\mu K$ arcmin$^2$ with covariance matrix shown in Fig.~\ref{fig:cor_diskring_ksz_uniformweight_bootstrap}.
Larger CAP filters are highly correlated and no new information is acquired when increasing the CAP apertures to $R=5$ arcmin.
Additionally, we converted the kSZ temperatures into integrated optical depth to Thomson scattering in the right vertical axis as $\tau_{\rm CAP} = T_{\rm kSZ}/T_{\rm CMB} \cdot c/\sigma_{v}^{\rm rec}$.
We evaluate the signal-to-noise ratio as in Eq.~\ref{eq:SNR_null}.
We use the Illustris $z=0.5$ curve as the reference ``model'' for the $S/N$, leading to $S/N = 9.8$.

\subsection{Simulation predictions: Illustris and IllustrisTNG}
\label{sec:sims}

We compare our results with simulations following the methodology outlined in \cite{hadzhiyska2024evidencelargebaryonicfeedback}. 
This is possible because our spectroscopic LRG galaxies are selected based on their photometric counterparts.
The spectroscopic LRG galaxies used in this work are expected to be a fair subsample of the photometric LRG galaxies, and thus share the same average gas profile \cite{bianchi2024characterizationdesifiberassignment}. 
Deviations from this due to fiber assignment incompleteness may occur in massive clusters due to fiber collisions, however we have masked clusters from our CMB map (with a completeness of 90$\%$ above $M_{500c} \sim 4 \cdot 10^{14} M_{\odot}$ \cite{Hilton_2021}).

Same as \cite{hadzhiyska2024evidencelargebaryonicfeedback}, we use fixed-redshift boxes from Illustris\footnote{\url{https://www.illustris-project.org/data/}} \cite{Nelson_2015, Vogelsberger_2014} and IllustrisTNG\footnote{\url{https://www.tng-project.org/data/}} (The Next Generation) \cite{nelson2021illustristngsimulationspublicdata}, and select galaxies using an abundance-matching approach. This approach ensures, first, that the number density of photometric LRG-type galaxies matches, and second, that the mean halo mass corresponds to the values inferred from an HOD analysis of DESI LRG galaxies at redshifts $0.4 < z < 0.6$ ($10^{13.4} M_{\odot}/h$) and $0.6 < z < 0.8$ ($10^{13.2} M_{\odot}/h$) \cite{yuan2023desi}.
These galaxies are then stacked at their respective locations on the simulated ACT maps.

\begin{figure}[]
    \centering
    \includegraphics[width=0.45\textwidth]{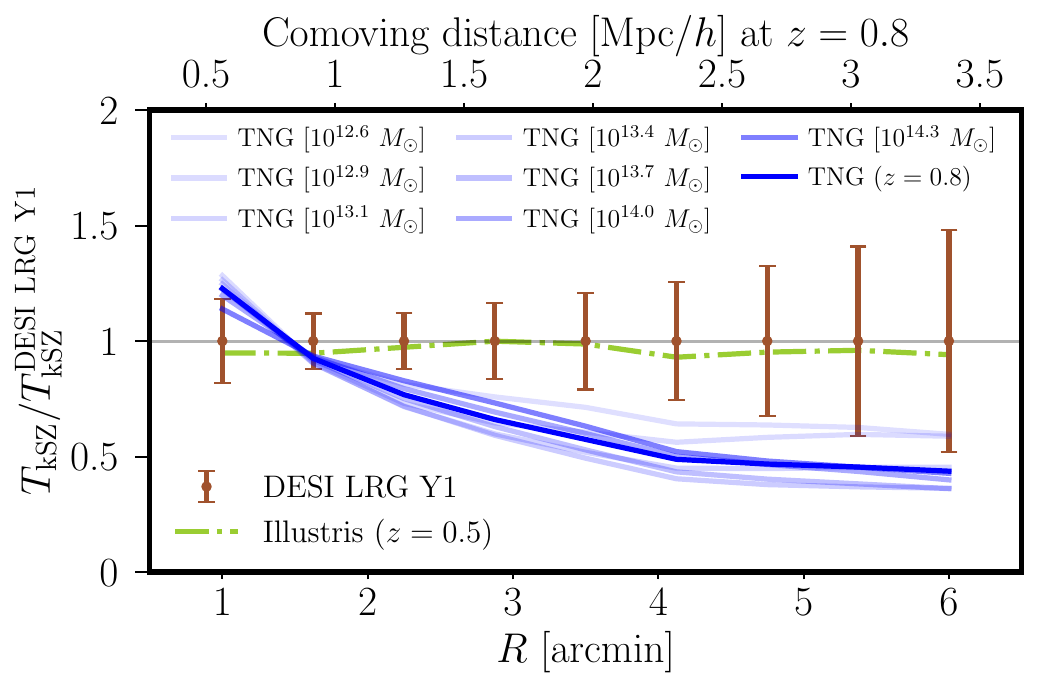}
    \caption{
     As the lower panel from Fig.~\ref{fig:kSZ_signal}, but the profiles are divided by our fiducial result (DESI LRG Y1).
     Additionally, a subset of IllustrisTNG is divided into four halo mass bins spanning $10^{12.6}$ to $10^{14.3}$ $M_{\odot}$. 
     These values are substantially larger than five times the uncertainty in the mean host halo mass of the DESI LRG galaxies \cite{yuan2023desi}.
     This analysis further indicates that the shapes of the IllustrisTNG profiles do not vary significantly across different mean masses—in other words, they do not adjust in a way that better aligns with our results.
     }
     \label{fig:shape_analysis}
\end{figure}

In particular, we study the beam-convolved kSZ CAP profiles from the Illustris and IllustrisTNG simulations at $z = 0.8$, which is close to the median redshift of the DESI LRG Y1 galaxies (Med($z$) = 0.75).
Specifically, we focus on the higher-resolution simulations: Illustris-1 (with volume $\sim$ (100 cMpc)$^3$) and IllustrisTNG300-1 (with volume $\sim$ (300 cMpc)$^3$), respectively.
We include the dark matter profile from the IllustrisTNG simulation around the same halos at $z=0.8$.
Due to the relatively small volume of simulations like Illustris, the most massive halos in our sample are not adequately represented. 
As expected, this can have a strong effect on the amplitude of the simulated profiles, since $\delta T_{\rm kSZ} \propto \tau \propto M_{\rm halo}$ (see Eq. \ref{eq:kSZ_estimator}). 
However, we have checked that this has a smaller effect on the profile shape (Fig.~\ref{fig:shape_analysis}).

To address this, we explore two approaches: 
The first approach fixes the outer region of the simulation so that it overlaps with the last data point of the data, where we expect all the halo gas to be fully encompassed (upper panel of Fig.~\ref{fig:kSZ_signal}).
The second approach completely disregards amplitude information in the data and simulation by fitting for a free amplitude parameter in the simulated profile (lower panel of Fig.~\ref{fig:kSZ_signal}).
This second approach is conservative in that any model ruled out even when the amplitude is allowed to vary freely would be even more ruled out otherwise.
The corresponding rescaling factors, $\chi^2_{\rm best-fit}$ and probability to exceed (PTE) are displayed in Table~\ref{tab:aperture_matched_filter}.
We leave to future work a more careful analysis, with simulations large enough to sample the highest-mass halos in our catalog, such that profile amplitude and shape can both be used.

\begin{table}
	\centering
	\begin{tabular}{|c|c|c|c|}
		\hline
            \multicolumn{4}{|c|}{\textbf{Matching the largest aperture}} \\
            \hline
            Simulation & Rescaling factor & $\chi^2_{\rm best-fit}$ & PTE \\
            \hline
            Illustris ($z=0.5$) & 1.80 & 9.13 & 0.331 \\
            \hline
            Illustris ($z=0.8$) & 1.89 & 40.56 & $\sim 10^{-6}$ \\
            \hline
            IllustrisTNG ($z=0.8$) & 0.86 & 242.04 & $\sim 10^{-48}$ \\
            \hline
            DM TNG ($z=0.8$) & 0.75 & 542.62 & $\sim 10^{-112}$ \\
            \hline
            \hline
            \multicolumn{4}{|c|}{\textbf{Freeing the simulated profile amplitude}} \\
            \hline
            Simulation & Rescaling factor & $\chi^2_{\rm best-fit}$ & PTE \\
            \hline
            Illustris ($z=0.5$) & 1.68 & 8.65 & 0.372 \\
            \hline
            Illustris ($z=0.8$) & 1.23 & 13.68 & 0.091 \\
            \hline
            IllustrisTNG ($z=0.8$) & 0.33 & 15.77 & 0.046 \\
            \hline
            DM TNG ($z=0.8$) & 0.21 & 21.91 &  0.005 \\
            \hline
	\end{tabular}
     \caption{
    The specific rescaling factors, chi-squared statistic $\chi^2_{\rm best-fit}$ and probability to exceed (PTE), from Fig.~\ref{fig:kSZ_signal}.
    }
\label{tab:aperture_matched_filter}
\end{table}
%

\subsubsection{Approach 1: matching the largest aperture}
\label{sec:approach1}

In the first simple approach, we rescale the simulated profiles by a multiplicative amplitude parameter, such that they match the measured data point at the largest aperture, as shown in the upper panel of Fig.~\ref{fig:kSZ_signal}.
We have included the corresponding rescaling factors and other statistics on the upper panel of Table~\ref{tab:aperture_matched_filter}, from which we could infer similar conclusions as the ones drawn in \cite{hadzhiyska2024evidencelargebaryonicfeedback} regarding the preference for large feedback and ruling out the dark matter and IllustrisTNG profiles.

In principle, this rescaling of the profile amplitudes seems justified since the largest aperture filters should encompass all the gas bound to halos, and thus they should match.
However, the following caveats are in order with this approach.
First and most concerning, the largest radii of the CAP also have the largest noise due to the primary CMB.
Thus, a noise fluctuation in this largest aperture data point could produce a large shift in the profile amplitude, leading to more or less agreement at smaller apertures.
Second and likely less concerning, the two-halo term becomes relevant at large apertures (Fig.~19 in \cite{Schaan2021}), such that matching at large apertures would also be enforcing a constant bias for the gas.
This is probably a reasonable assumption, but would require more work to validate.
We thus propose a second, more conservative approach below.

\subsubsection{Approach 2: freeing the simulated profile amplitude}
\label{sec:approach2}

In this subsection, we take a more conservative approach, entirely discarding the profile amplitude information in the data and the simulation.
Indeed, we allow for a free amplitude parameter multiplying each of the simulated profiles, and fit them to minimize the $\chi^2$ between data and model.
The standard linear fitting procedure is explained in App.~\ref{sec:matched_filter_aperture}.
The resulting simulated profiles with best-fit amplitudes are shown in the lower panel of Fig.~\ref{fig:kSZ_signal} and the corresponding best fit amplitude and goodness of fit in the lower panel of Table~\ref{tab:aperture_matched_filter}.

We have also verified that obtaining the kSZ profile from simulations with masses varying around our mean -and hence, different amplitudes- has a small effect on the profile shape (see Fig.~\ref{fig:shape_analysis}) and, therefore, on the conclusions drawn from Fig.\ref{fig:kSZ_signal}.
This motivates the approach of freeing the amplitude while preserving the shape of the profiles.

With this conservative approach, the hypothesis that the measured gas profile would follow the simulated dark matter profile is still disfavored with PTE = 0.005,
in line with previous studies 
\cite{Schaan2021, hadzhiyska2024evidencelargebaryonicfeedback}.
The dark matter profile is too concentrated to provide a good fit.
IllustrisTNG (PTE = 0.046) predicts a less extended gas profile compared to Illustris (PTE = 0.091) at $z=0.8$, but none is significantly ruled out in this conservative approach.
However, for IllustrisTNG, this requires a very small best-fit amplitude of $0.33$, meaning that the mean halo mass in the IllustrisTNG sample would have to be underestimated by a factor 3.
This is unlikely, and is instead likely a sign that this simulated profile with low feedback is actually too concentrated.
A more thorough analysis will be required to conclude on this, and we leave it to future work.

Intriguingly, the shape of the Illustris profile at $z=0.5$ appears to be the best match to our data (at $z_\text{median} = 0.75$, PTE = 0.372).
The treatment of stellar and AGN feedback is known to cause discrepancies with observations, notably in the X-ray regime and when comparing model predictions to observed tSZ profiles (see Section 6.1 of \cite{Nelson_2015}).
A strong deviation in the tSZ signal is also observed in IllustrisTNG \cite{Moser2022, https://doi.org/10.48550/arxiv.2307.10919}, which aimed to address additional mismatches by incorporating new physics and numerical improvements relative to its predecessor (see Table 2 of \cite{nelson2021illustristngsimulationspublicdata}).
We hope future analyses of our measurements can shed more light on the reason for the preference for the $z=0.5$ Illustris gas profile.

A thorough comparison of our measurements to simulation will need to accurately match the simulated galaxy sample in hydrodynamical simulation with the actual one. This includes the correct distribution of host halo masses, satellite fraction (including the corresponding miscentering), and any other selection criterion (such as color cuts and masking of massive clusters) applied to the data \cite{Moser2021, Moser2022, mccarthy2024flamingocombiningkineticsz}.
In particular, \cite{mccarthy2024flamingocombiningkineticsz} showed that satellites could have an important effect on the simulated kSZ signal.
In the case of DESI LRG, \cite{hadzhiyska2024evidencelargebaryonicfeedback} varied the satellite fraction from 0$\%$ to 30$\%$ and found that the corresponding profile shape was not significantly altered.
We leave this thorough investigation to future work.
%

\subsection{Agreement with previous spectroscopic measurements}

In \cite{Schaan2021}, the kSZ signal was detected using spectroscopic galaxies from CMASS and a previous ACT map (DR5). 
A Gaussian CAP profile provided a reasonable fit, especially given the reported uncertainty, however for DESI LRG Y1, this is no longer true, as the data, especially the CMB map and masks, had improved substantially.
For a detailed comparison between the CMASS and DESI samples, see App.~\ref{sec:comparison_cmass}.

\subsection{Dependence on galaxy properties}
\label{sec:ksz_dependence}

We divide our galaxy sample based on redshift, stellar mass, and absolute magnitude and derive their kSZ CAP profiles, as illustrated in Figs.~\ref{fig:kSZ_per_bin_LRG}, \ref{fig:kSZ_per_mass_bin_massweight}, and \ref{fig:Abs_mag_kSZ}, respectively. 

\begin{figure}
     \centering
     \begin{subfigure}[b]{0.42\textwidth}
         \centering
         \includegraphics[width=0.99\textwidth]{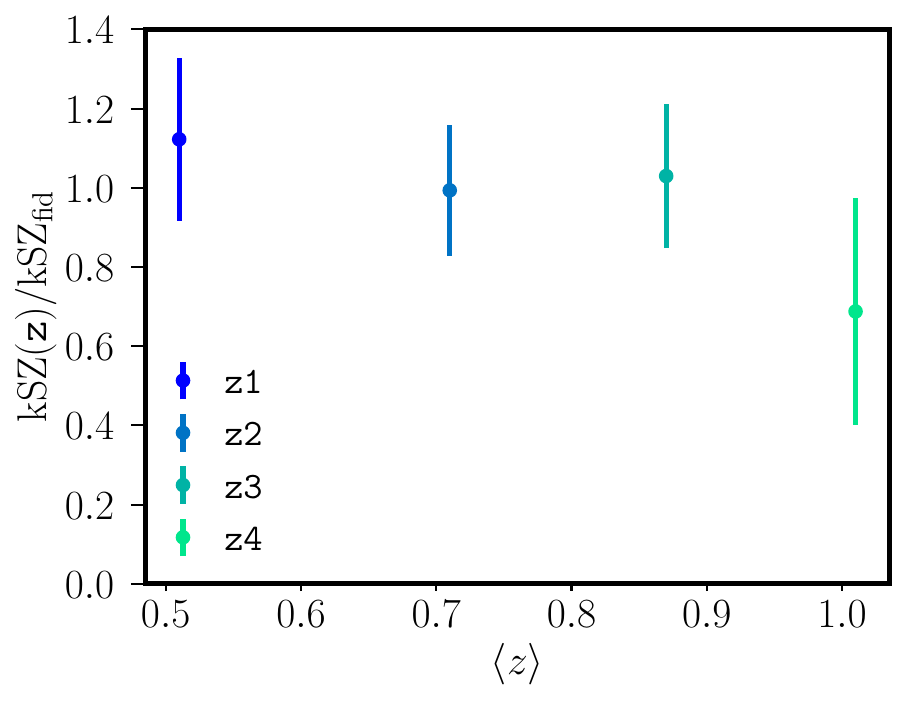}
     \end{subfigure}
     \hfill
     \begin{subfigure}[b]{0.42\textwidth}
         \centering         \includegraphics[width=0.99\textwidth]{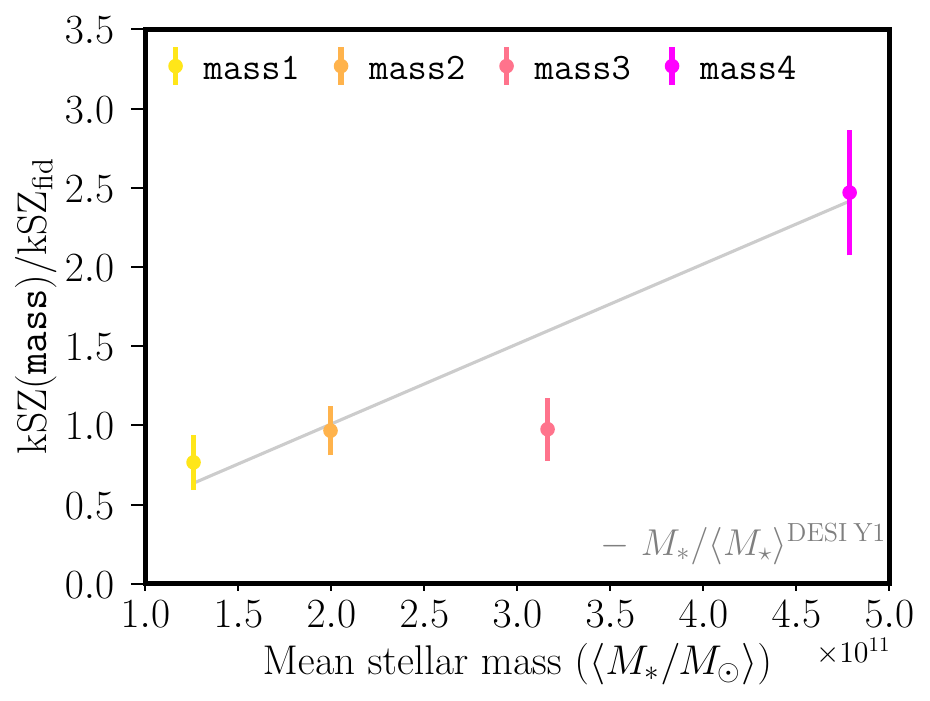}
     \end{subfigure}
     \hfill
     \begin{subfigure}[b]{0.42\textwidth}
         \centering         \includegraphics[width=0.99\textwidth]{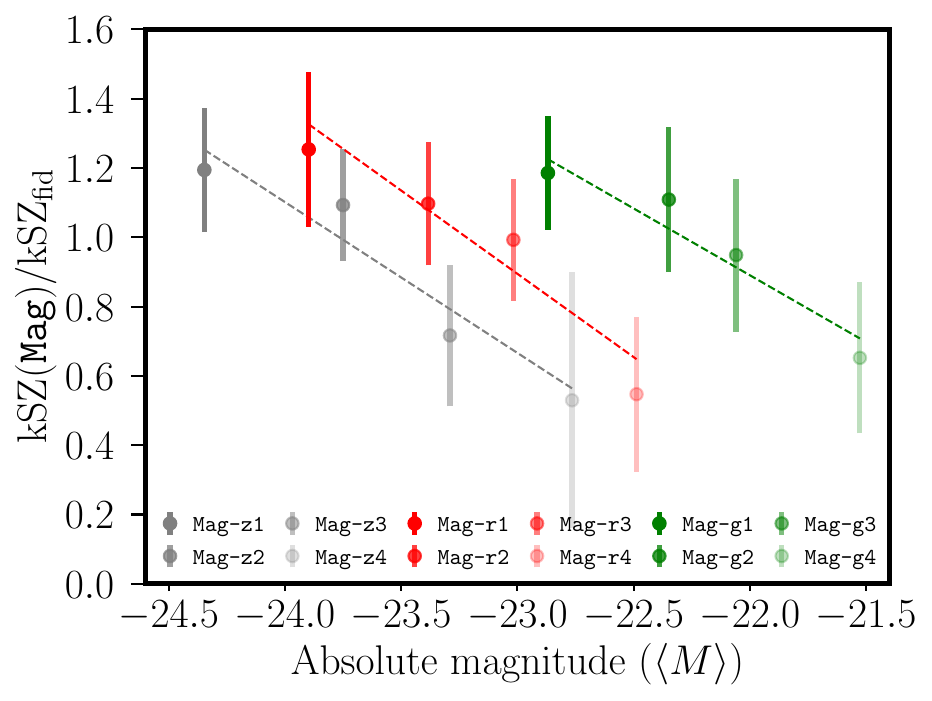}
     \end{subfigure}
     \caption{
     Fractional amplitudes of the stacked kSZ profiles for different variables and their corresponding bins (from Figs. \ref{fig:kSZ_per_bin_LRG}, \ref{fig:kSZ_per_mass_bin_massweight} and \ref{fig:Abs_mag_kSZ}) with respect to the fiducial one from Fig.~\ref{fig:kSZ_signal}.
     \textit{Upper panel:}
     For higher redshift, we find no trend on kSZ signals compared to our large error bars.
     \textit{Central panel:}
     The highest mass bin has an amplitude $\sim3$ times larger than the one observed in the whole fiducial analysis.
     This result matches the mass dependence \cite{hadzhiyska2024evidencelargebaryonicfeedback} found when also using LRG galaxies.
     \textit{Bottom panel:}
     We examine three optical bands: $z$, $r$, and $g$, matched with their respective associated colors. 
     In all of the absolute magnitude cases, the less luminous a galaxy is, the smaller the measured amplitude of $T_{\rm kSZ}$, as shown by the weighted least squares fit (dashed lines) and their negative slopes.
     }
     \label{fig:amplitudes}
\end{figure}

We compare the measured kSZ profile in each galaxy subsample (e.g., specific redshift bin or magnitude bin) to the fiducial kSZ profile for the whole sample.
Given the limited SNR in each subsample, we compress this comparison to a single amplitude parameter: the relative amplitude between the kSZ profile in the subsample versus the whole sample (displayed in Fig. \ref{fig:kSZ_signal}).
This process is detailed in App.~\ref{sec:matched_filter_aperture}.
Our results show that the kSZ signal evolves with stellar mass and absolute magnitude (Fig.~\ref{fig:amplitudes}). 
The corresponding summary statistics are provided in Table~\ref{tab:result_reports}.

\begin{table*}
	\centering
	\begin{tabular}{|c|c|c|c|c|c|c|c|c|}
		\hline
            Bin & $\langle z \rangle$ & Med($z$) & $10^{11} \hspace{0.05 cm} \langle M_{\star} \rangle$/$M_{\odot}$ & $N$ & $\chi^2_{\rm null}$ & $S/N$ & kSZ/kSZ$_{\rm fid}$ & $\sigma_{\rm kSZ}/\sigma_{\rm kSZ_{\rm fid}}$ \\
            \hline
            \textbf{Full sample} & \textbf{0.74} & \textbf{0.75} & $\mathbf{2.2}$ & \textbf{825,283} & \textbf{105.9} & \textbf{9.8} & \textbf{1.0} & \textbf{0.0} \\
            \hline
            \texttt{z1} & 0.51 & 0.51 & $2.4$ & 195,877 & 36.5 & 5.6 & 1.12 & 0.21 \\
            \texttt{z2} & 0.71 & 0.71 & $2.3$ & 297,440 & 46.6 & 5.1 & 0.99 & 0.16 \\
            \texttt{z3} & 0.87 & 0.87 & $2.0$ & 235,620 & 42.8 & 5.8 & 1.03 & 0.18 \\
            \texttt{z4} & 1.01 & 1.01 & $2.1$ &  96,346 & 9.6  & 2.3 & 0.69 & 0.29 \\
            \hline
            \texttt{mass1} & 0.76 & 0.79 & $1.2$ & 244,932 & 20.6 & 4.3 & 0.77 & 0.17 \\
            \texttt{mass2} & 0.75 & 0.76 & $2.0$ & 320,914 & 44.4 & 6.0 & 0.97 & 0.16 \\
            \texttt{mass3} & 0.71 & 0.70 & $3.0$ & 194,037 & 26.7 & 4.5 & 0.98 & 0.20 \\
            \texttt{mass4} & 0.69 & 0.67 & $5.1$ & 53,997  & 47.7 & 6.2 & 2.47 & 0.40 \\
            \hline
            \texttt{Mag-g1} & 0.80 & 0.82 & $2.9$ & 299,592 & 55.2 & 7.1 & 1.19 & 0.16 \\
            \texttt{Mag-g2} & 0.73 & 0.74 & $2.2$ & 189,580 & 40.0 & 5.1 & 1.11 & 0.21 \\
            \texttt{Mag-g3} & 0.70 & 0.71 & $1.9$ & 164,457 & 24.2 & 4.2 & 0.95 & 0.22 \\
            \texttt{Mag-g4} & 0.68 & 0.70 & $1.5$ & 168,848 & 15.3 & 2.5 & 0.65 & 0.22 \\ 
            \hline            
            \texttt{Mag-r1} & 0.81 & 0.84 & $3.5$ & 164,927 & 43.7 & 5.6 & 1.25 & 0.22 \\
            \texttt{Mag-r2} & 0.75 & 0.76 & $2.4$ & 250,337 & 42.1 & 5.9 & 1.10 & 0.18 \\
            \texttt{Mag-r3} & 0.72 & 0.74 & $1.8$ & 250,628 & 42.6 & 5.6 & 0.99 & 0.18 \\
            \texttt{Mag-r4} & 0.68 & 0.70 & $1.3$ & 159,688 & 11.6 & 2.0 & 0.55 & 0.22 \\
            \hline 
            \texttt{Mag-z1} & 0.77 & 0.79 & $3.4$ & 257,941 & 50.9 & 6.7 & 1.19 & 0.18 \\
            \texttt{Mag-z2} & 0.74 & 0.75 & $2.2$ & 315,034 & 52.3 & 6.8 & 1.09 & 0.16 \\
            \texttt{Mag-z3} & 0.74 & 0.76 & $1.6$ & 194,551 & 22.9 & 2.3 & 0.72 & 0.20 \\
            \texttt{Mag-z4} & 0.69 & 0.73 & $1.0$ &  58,054 & 12.6 & 1.1 & 0.53 & 0.37 \\
            \hline
	\end{tabular}
     \caption{
     Statistics of the DESI LRG Y1 sample kSZ and across various bins when splitting it by redshift (denoted as \texttt{z} followed by the bin number), stellar mass (denoted as \texttt{mass} followed by the bin number), and absolute magnitude for the three optical bands (denoted as \texttt{Mag-} followed by the color and bin number). 
     For a given bin and total sample, we report the mean redshift $\langle z \rangle$, the median redshift Med$(z)$, the mean stellar mass $\langle M_{\star} \rangle$ in units of solar masses ($M_{\odot}$) and the number of galaxies $N$.
     We also include $\chi^2_{\rm null}$ from Eq.~\ref{eq:chi_squared} with nine degrees of freedom, which quantifies the rejection of the null hypothesis, and the corresponding $S/N$ (Eqs. \ref{eq:SNR_null}).
     Finally, we included the amplitudes and uncertainties shown in Fig.~\ref{fig:amplitudes}.
     For the complete sample, we find $S/N = 9.8$.
    }
\label{tab:result_reports}
\end{table*}

\subsubsection{Absence of clear redshift evolution}
\label{sec:redshift_dependence}

The baryonic feedback processes which impact galaxy formation and evolution are expected to evolve in time \cite{martinalvarez2024stirringcosmicpotblack}.
It is thus natural to expect the gas profiles around galaxies to evolve with time as well.
Measuring the gas distribution through the kSZ as a function of redshift can give us clues about this puzzle.

We divide the DESI LRG Y1 sample into four redshift bins, as listed in Sec.~\ref{sec:desi_data}, and in Table~\ref{tab:result_reports}, we report the summary statistics resulting in a significant detection in the first three redshift bins.
For \texttt{z4}, the lower $S/N$ is partially explained by the low number of LRG galaxies with $z \in [0.95, 1.10]$.

\begin{figure}
    \centering
    \includegraphics[width=0.48\textwidth]{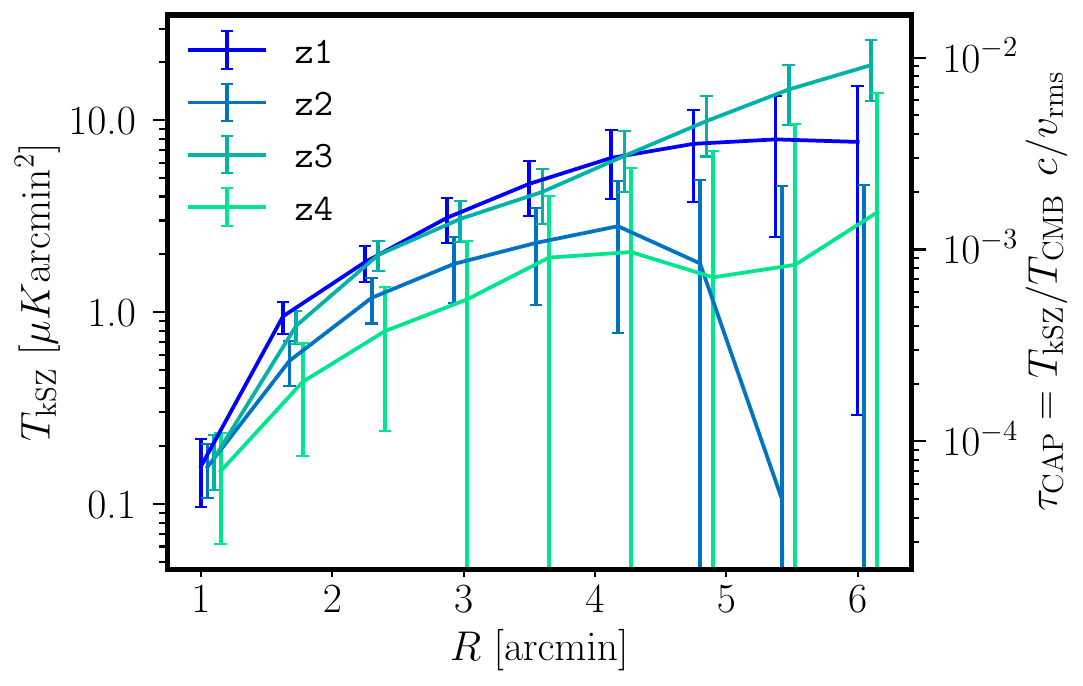}
    \caption{
    Mean stacked kSZ profiles for the different redshift bins.
    There is no clear trend of the profiles, confirming the results obtained by \cite{hadzhiyska2024evidencelargebaryonicfeedback}. 
    Additionally, the number of galaxies in the redshift bins is not evenly distributed: \texttt{z4} reports the lowest $S/N$ $= 2.3$ with 96,346 galaxies, which e.g.~corresponds to only $\sim 1/3$ of the galaxies of bin \texttt{z2}.
    }
    \label{fig:kSZ_per_bin_LRG}
\end{figure}

\begin{figure}
    \centering
    \includegraphics[width=0.48\textwidth]{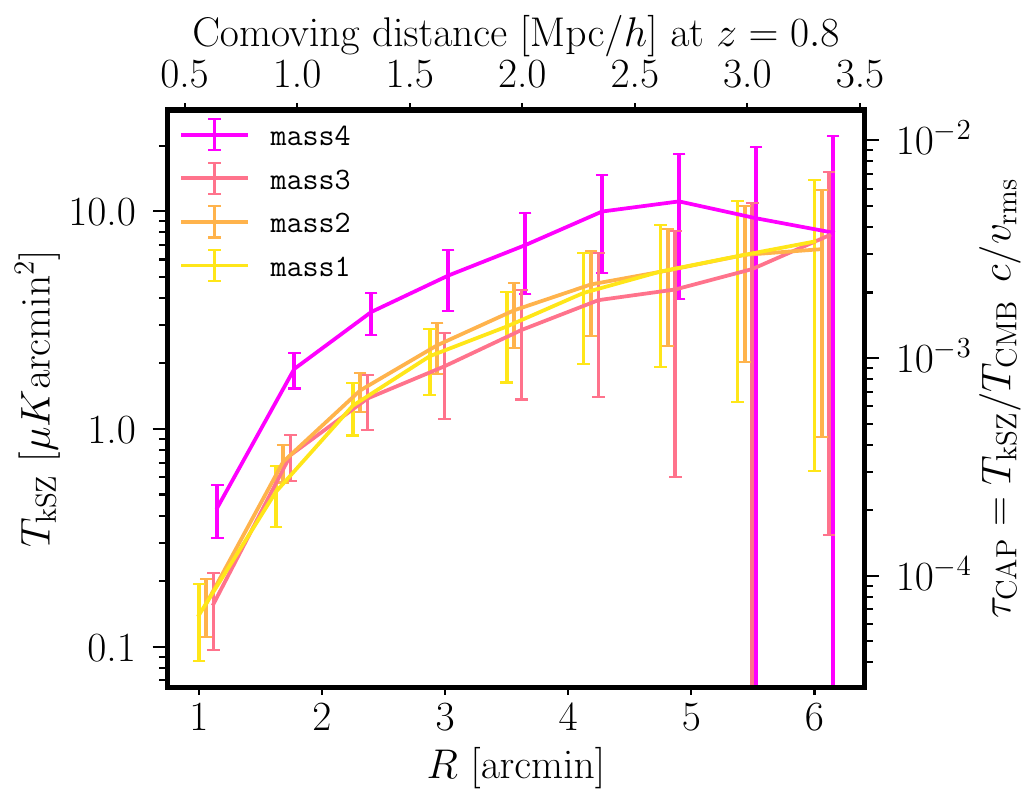}
    \caption{
    Similar to Fig.~\ref{fig:kSZ_per_bin_LRG}, the kSZ stacked profiles for the different stellar mass bins, denoted by \texttt{mass}.
    For \texttt{mass4}, there is a clear increment on the amplitude, similar to what \cite{hadzhiyska2024evidencelargebaryonicfeedback} found.
    At larger scales, the errors are wider and more correlated, thus the data at large $R$ provides little new information beyond that from the smaller scales.
    }
    \label{fig:kSZ_per_mass_bin_massweight}
\end{figure}

In Fig.~\ref{fig:kSZ_per_bin_LRG}, we present the corresponding $T_{\rm kSZ}$ CAP profiles.
The large noise makes it difficult to detect any evolution of the profile shape with redshift. 
An additional factor that could lead to variations in the kSZ profiles within the same sample is that, at different redshifts, a fixed CAP filter radius corresponds to different proper distances and therefore probes the gas distribution over different physical scales. 
Thus, at a fixed CAP radius, and in the absence of redshift-dependent feedback, we would inherently be probing a smaller fraction of the gas.
Although the large error bars in our current results prevent this effect from being visually apparent, it will become important to account for with upcoming data.
We compare the fractional kSZ amplitudes in all four redshift bins in the upper panel from Fig.~\ref{fig:amplitudes} and see no significant tendency with redshift.
We also measured the $\chi^2$ of the difference of paired profiles at different $z$, and confirm that the data are not constrained enough to elucidate a clear trend\footnote{We quantify this by calculating the maximum $\chi^2$ of the difference of paired profiles at different $z$. This corresponds to $\chi^2 = $19.47 for 18 degrees of freedom. The probability to exceed (PTE) is 23$\%$.}.

With a greater number of galaxies, it may be possible to differentiate between astrophysical effects that could enhance the feedback from galaxies at later epochs, thereby extending the spatial distribution of halo gas to larger angular scales.
A parameter to further explore is the satellite fraction as a function of redshift and mass, which we are fixing for in this work.
A parameter that we do not expect to vary substantially across redshift is the correlation coefficient $r$, as long as the galaxy sample is dense enough to recover the linear velocity field through Eq.~\ref{eq:continuity_eq} \cite{Ried_Guachalla_2024}.
As shown in Table~I of \cite{Hadzhiyska2023}, this parameter is robust to a variety of redshift distributions.
Furthermore, as shown in Fig. 7 of \cite{Ried_Guachalla_2024}, varying the satellite fraction between 10\% and 30\% does not substantially change $r$ (see the reconstruction of halo velocities, in dark green).
So although the satellite fraction of the sample may evolve as a function of redshift or stellar mass, which can affect the mean gas profile in interesting ways, it should not bias our estimate of $r$.

Finally, we acknowledge that additional factors could be causing a time evolution other than the evolution of feedback: for example, the galaxy selection function is certainly changing with redshift. 
This will need to be explored in future work.

\subsubsection{Strong stellar mass dependence at high masses}
\label{sec:mass_dependence}

The modalities of feedback (from supernovae or active galactic nuclei) as well as their amplitudes should depend on the properties of the galaxies considered. 
By measuring the gas profiles around galaxies as a function of stellar mass, we may therefore uncover valuable clues about the way feedback in galaxy formation works.

Using the DESI calibrated photometric stellar masses from \cite{Zhou_2023}, we split the sample into four mass bins as shown in Table~\ref{tab:result_reports}.
Under a similar argument as in \cite{hadzhiyska2024evidencelargebaryonicfeedback}, we split the sample maximizing the $S/N$ of individual bins (see Sec. \ref{sec:desi_data} for a detailed explanation) and find that higher stellar masses imply larger measured kSZ:
In Fig.~\ref{fig:kSZ_per_mass_bin_massweight}, we show the kSZ CAP aperture profiles, finding a considerably larger amplitude for the highest mass bin. 
The same tendency that was observed in Fig. 4 from \cite{hadzhiyska2024evidencelargebaryonicfeedback}.

At the cluster scale, the gas mass is expected to scale linearly with halo mass. 
However, at smaller scales, such as in galaxy groups, this relationship may be less pronounced. 
As a result, the stellar mass could help determine whether the measured kSZ effect follows a particular trend.
The central panel of Fig.~\ref{fig:amplitudes} shows that a linear model closely approximates the measurements, where $\langle M_{\star} \rangle ^{\rm DESI \hspace{0.1 cm} Y1}$ corresponds to the total mean stellar mass.
An important caveat is the unknown systematic or statistical error in the stellar mass estimates from photometry, which we use here.
In order to confirm the observed trend, it would be valuable in the future to additionally use more spectroscopic information to further calibrate the stellar masses.
If this tendency is still observed, it could become valuable input for simulations to tune their feedback models for varying stellar masses of galaxies.

\begin{figure}
    \centering
    \includegraphics[width=0.48\textwidth]{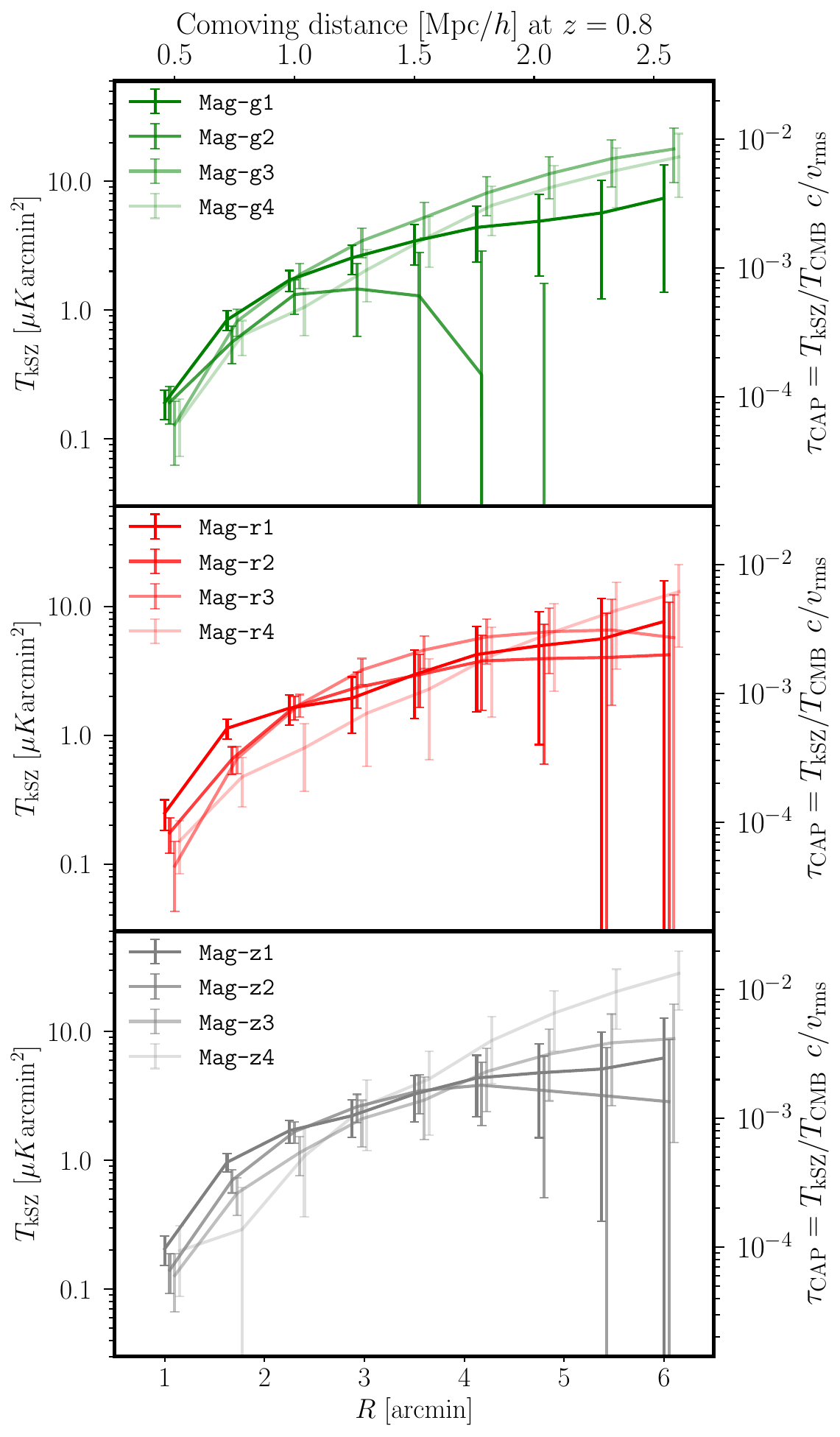}
    \caption{
    Similar to Figs. \ref{fig:kSZ_per_bin_LRG} and \ref{fig:kSZ_per_mass_bin_massweight}, the kSZ stacked CAP profiles for each absolute magnitudes bin, denoted \texttt{Mag-n}, with \texttt{n} corresponding to one of the photometric bands.
    As in the redshift study case, the dependence on luminosity is not visually clear, however the bottom panel of Fig.~\ref{fig:amplitudes} shows a increasing trend with absolute magnitude for all the cases.
    }
    \label{fig:Abs_mag_kSZ}
\end{figure}
%


\subsubsection{Clear dependence on absolute magnitude}
\label{sec:abs_mag_dependence}

\begin{figure*}
    \centering
    \includegraphics[width=0.95\textwidth]{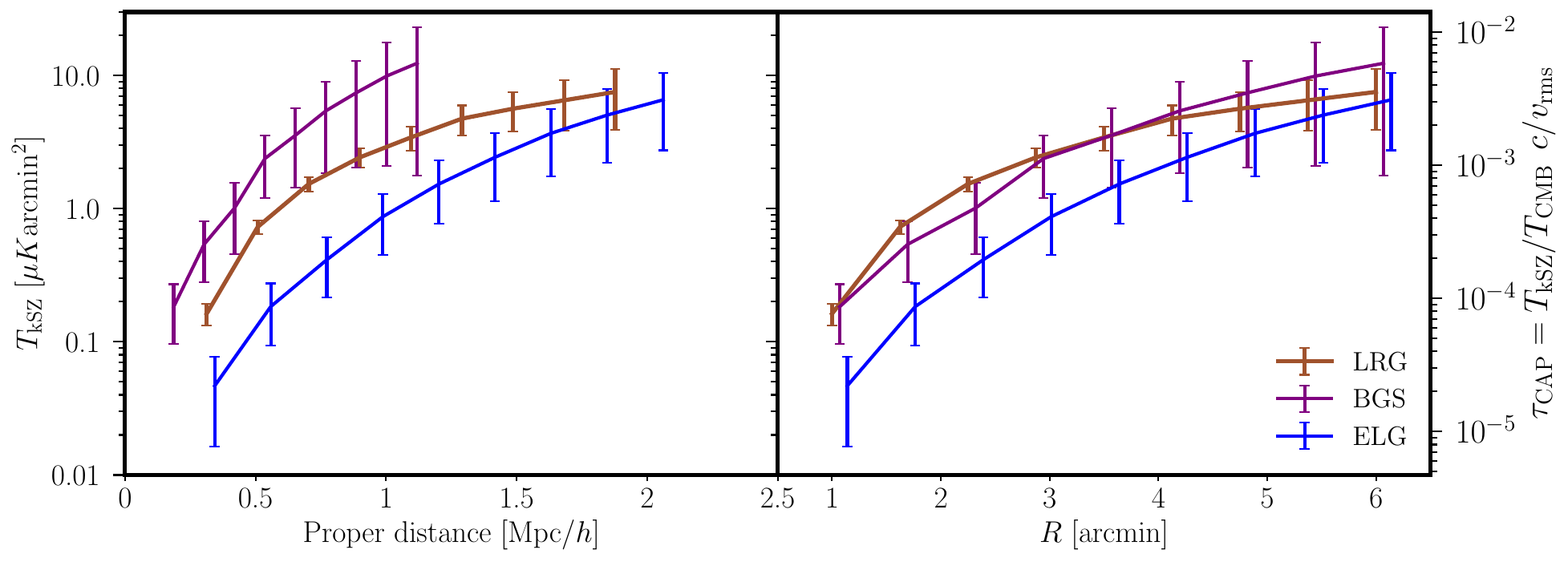}
    \caption{
    Comparison between the kSZ stacked CAP profiles obtained when using the different galaxy samples from DESI Y1: BGS (in purple) and ELG (in blue), both compared to the LRG main result.
    In the \textit{left panel}, we plot the profiles as a function of the proper distance using the mean redshift of each sample, while for the \textit{right panel}, we use the observed angular scale.
    Within the noise, the observed BGS profile shows both a shape and amplitude consistent with the LRG profile. 
    One would have na\"ively expected the BGS sample to have a smaller kSZ amplitude and different gas profile, since they are typically hosted by lower mass halos.
    However, our kSZ measurement includes only the brighter subset of BGS galaxies used in the DESI Y1 BAO reconstruction, leading to a higher mean mass compared to the full sample (see \cite{desicollaboration2024desi2024iisample} for more details).    
    The ELG sample, which includes a significant fraction of satellite galaxies, exhibits two main combined effects: a more extended profile due to satellite contamination and larger uncertainties arising from the virial motions of the satellite galaxies.
    }
    \label{fig:lrg_bgs_elgs_y1}
\end{figure*}

Another galaxy property expected to correlate with gas mass is galaxy absolute magnitude, since brighter galaxies tend to live in more massive dark matter halos, which thus contain more mass.
For this reason, we also explore the dependence of kSZ signal on the galaxies' absolute luminosities in the $g$, $r$, and $z$ bands.

In Fig.~\ref{fig:Abs_mag_kSZ}, we show the kSZ CAP aperture profiles for the three bands; however, the large errors, especially for apertures $R>4$ arcmin, prevent us from distinguishing between the shapes of the profiles. 
A larger sample could help to better constrain these profiles, as random errors would decrease substantially with future data releases, as discussed in Sec. \ref{sec:forecast}.
In the bottom panel of Fig.~\ref{fig:amplitudes}, we observe an increasing fractional amplitude of the kSZ signal with absolute magnitude, consistent with a weighted least squares fit that shows negative slopes: -0.44, -0.48, and -0.39, for $z$, $r$, and $g$, respectively.
This confirms that gaseous halos are more massive around more luminous galaxies.


\subsection{Comparison with bright and emission-line galaxy samples from DESI Y1}
\label{sec:comparison_elgs_bgs}

We have mainly studied LRG due to their clustering properties and ease of detection, however, understanding the physics of galaxy formation and evolution is not limited to a single galaxy type, but all possible ones.
In this section, we calculate the kSZ CAP profiles of two additional DESI samples overlapping with the ACT map.
The first one is the Bright Galaxy Sample (BGS), which consists of 95,934 galaxies with a roughly constant number density in the range $0.1 \le z \le 0.4$ \cite{Hahn_2023}.
The division between these galaxies and the LRG sample mainly relies on the fact that BGS probe the epoch when dark energy becomes dominant; however, the BGS sample contains a wide range of galaxy types.
The second one is the Emissison-Line Galaxy sample (ELG), which are OII emission-line galaxies that are generally active star-forming galaxies and are less clustered than the LRG ones \cite{Raichoor_2023}.
This sample consists of 1,000,998 galaxies spanning a large redshift range of $0.8 \le z \le 1.6$.
Both redshift distributions are shown in Fig.~1 of \cite{desicollaboration2024desi2024iisample}.
We use the velocity correlation coefficients $r$ from \cite{Hadzhiyska2023}: $r_{\rm BGS} = 0.64$ and $r_{\rm ELG} = 0.55$. 
Similar to our approach for the LRG, we use the BAO displacements and convert them into velocities, as explained in Sec.~\ref{sec:vel_rec}.
We leave a detailed modeling of these profiles to future work.

Retaining the same model as for our LRG kSZ measurement (Illustris gas profiles at $z=0.5$), we obtain $S/N$ = 2.3 and 2.1 for ELG and BGS galaxies, respectively.
Fitting these measurements with a more accurate model would lead to a higher $S/N$, and we leave that task to a future work as it would require an exploration of the BGS and ELG population in simulations.

In Fig.~\ref{fig:lrg_bgs_elgs_y1}, we show both CAP profiles of BGS and ELG next to the more precise measurement from the LRG sample.
To highlight the differences between the profiles, and their physical extension, we plot them as a function of proper distance and their observed angular scale.
We find that the amplitude of the kSZ profiles for LRG and BGS galaxies appear to match, similar to what was found in \cite{Hadzhiyska:2024ecq} with the photometric samples.
One would expect BGS galaxies to have a lower mean host halo mass \cite{Hahn_2023}, and therefore, a lower amplitude on their kSZ profile.
The DESI BGS Y1 sample, however, has been restricted to include only lower-magnitude objects \cite{desicollaboration2024desi2024iisample}. 
This magnitude cut significantly reduces the total number of galaxies (by approximately 90$\%$) and likely increases the mean host halo mass, bringing it into closer agreement with that of the LRG. 
Indeed, the stellar mass estimate from photometry corresponds to $\sim 3 \cdot 10^{11} M_{\odot}$, close to the one find for our sample ($2.2 \cdot 10^{11} M_{\odot}$).
For the ELG sample, the measured gas profile appears more extended compared to that of the LRG sample. 
This can be attributed to the larger fraction of satellites in the ELG sample, as compared to the LRG sample, when using the same halo-finding method (e.g., \cite{rocher2023desi, yuan2023desi}).
As shown in \cite{hadzhiyska2024evidencelargebaryonicfeedback, mccarthy2024flamingocombiningkineticsz}, a large satellite fraction affects the shape and amplitude of the kSZ profiles, as some gaseous halos may be double-counted and miscentered.
And therefore, the 2D profiles tend to be observed distorted and extended.
Additionally, the mean host halo mass of the ELG differs from that of the LRG (the HOD analysis from \cite{rocher2023desi, yuan2023desi} estimates $M_{\rm halo}^{\rm ELG} \sim 10^{12} M_{\odot}/h$ and $M_{\rm halo}^{\rm LRG} \sim 10^{13} M_{\odot}/h$, respectively).


\subsection{Consistency with photometric kSZ}
\label{sec:comparison_boryana}

Photometric surveys allow us to create large galaxy catalogs, but they have the drawback of missing important redshift information.
In contrast, spectroscopic surveys provide a smaller yet more precise 3D distribution of galaxies, which is why kSZ measurements have historically relied more on the latter.
A recent simulation-based study \cite{Ried_Guachalla_2024, Hadzhiyska2023} found that measuring the velocity stacking kSZ in photometric LRG-type galaxies is feasible, although it results in degraded $S/N$ which can be compensated by increasing the sample size (by approximately four times).
Consequently, \cite{hadzhiyska2024evidencelargebaryonicfeedback} measured the kSZ signal through velocity stacking using the DESI Legacy Imaging Survey \cite{Zhou_2023} and the ACT DR6 CMB, focusing specifically on the LRG targets for the complete DESI survey.
In this section, we compare both results.

\begin{figure}
    \centering
    \includegraphics[width=0.48\textwidth]{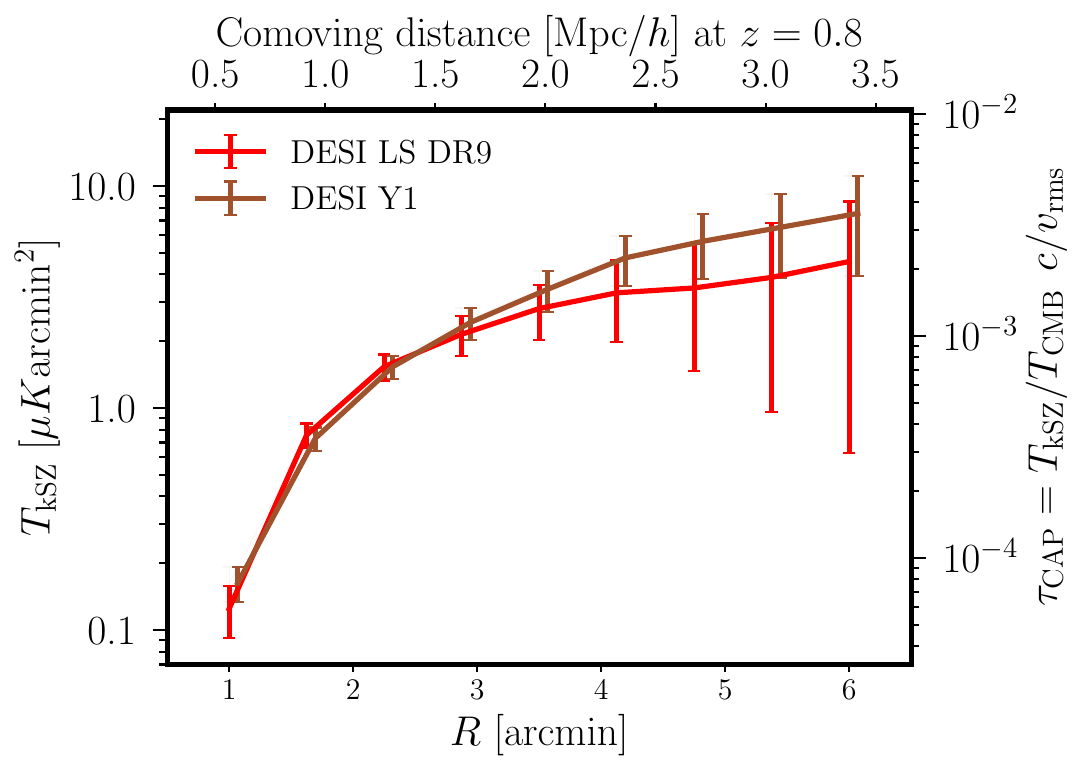}
    \caption{
    Comparison between the kSZ stacked CAP profiles obtained when using the photometric DESI LRG sample (from the DESI Legacy Imaging Surveys DR9 in red, subsample from what is used in \cite{hadzhiyska2024evidencelargebaryonicfeedback}) and its spectroscopic counterpart (from DESI Y1 in brown).
    The matching shapes for the smaller scales proves the robustness of the kSZ stacking method on both photometric and spectroscopic galaxies. 
    }
    \label{fig:photo_vs_spec}
\end{figure}

The photometric sample of LRG from the DESI Legacy Imaging Survey achieves a $\sigma_z/(1+z) \lesssim 0.02$.
It consists of two catalogs: the `Main LRG' and `Extended LRG', the latter having 2–3 times the DESI LRG density \cite{Zhou_2023_2}.
For an accurate comparison, we examine the `Main LRG' photometric sample from \cite{hadzhiyska2024evidencelargebaryonicfeedback}, which shares the same statistical properties as the spectroscopic sample in this paper, as it corresponds to the spectroscopic targets used by DESI.

For the DR9 sample, shown in the last row of Table 3 from \cite{hadzhiyska2024evidencelargebaryonicfeedback}, the kSZ is measured with $\chi_{\rm null} = 95.6$ an $S/N$ = 9.1 using a total of 3,118,161 galaxies.
When plotting the CAP profiles of the photometric and spectroscopic counterparts together, as shown in Fig.~\ref{fig:photo_vs_spec}, we find agreement in both the shapes and amplitudes. 
The reduced chi-squared statistic for the difference between these two measurements is $\Delta \chi_{\rm null}^{2}/{\rm dof} = (105.4 - 95.6)/9 = 9.8/9 \approx 1.1$, indicating that the profiles are consistent with each other (assuming their covariances are the same, after verifying the consistency of their covariances).

We note that we do not consider the highest $S/N$ result reported in \cite{hadzhiyska2024evidencelargebaryonicfeedback} because it includes additional corrections that are not directly transferable to our spectroscopic results.
These include the removal of outliers with large errors in their reconstructed velocities, as well as imposing a maximum cut on the estimated photo-$z$ error ($\sigma(z) < 0.05$).


\subsection{Current and future prospects: \textit{S}/\textit{N} forecasts}
\label{sec:forecast}

Since the first velocity stacking kSZ measurements, we have seen a steady growth of the $S/N$ with larger galactic catalogs.
Upcoming galaxy surveys, both spectroscopic and photometric, that overlap with CMB experiments will increment the $S/N$ further.
In this section, we study forecasts on the kSZ $S/N$.

\begin{figure*}
    \centering
    \includegraphics[width=0.85\textwidth]{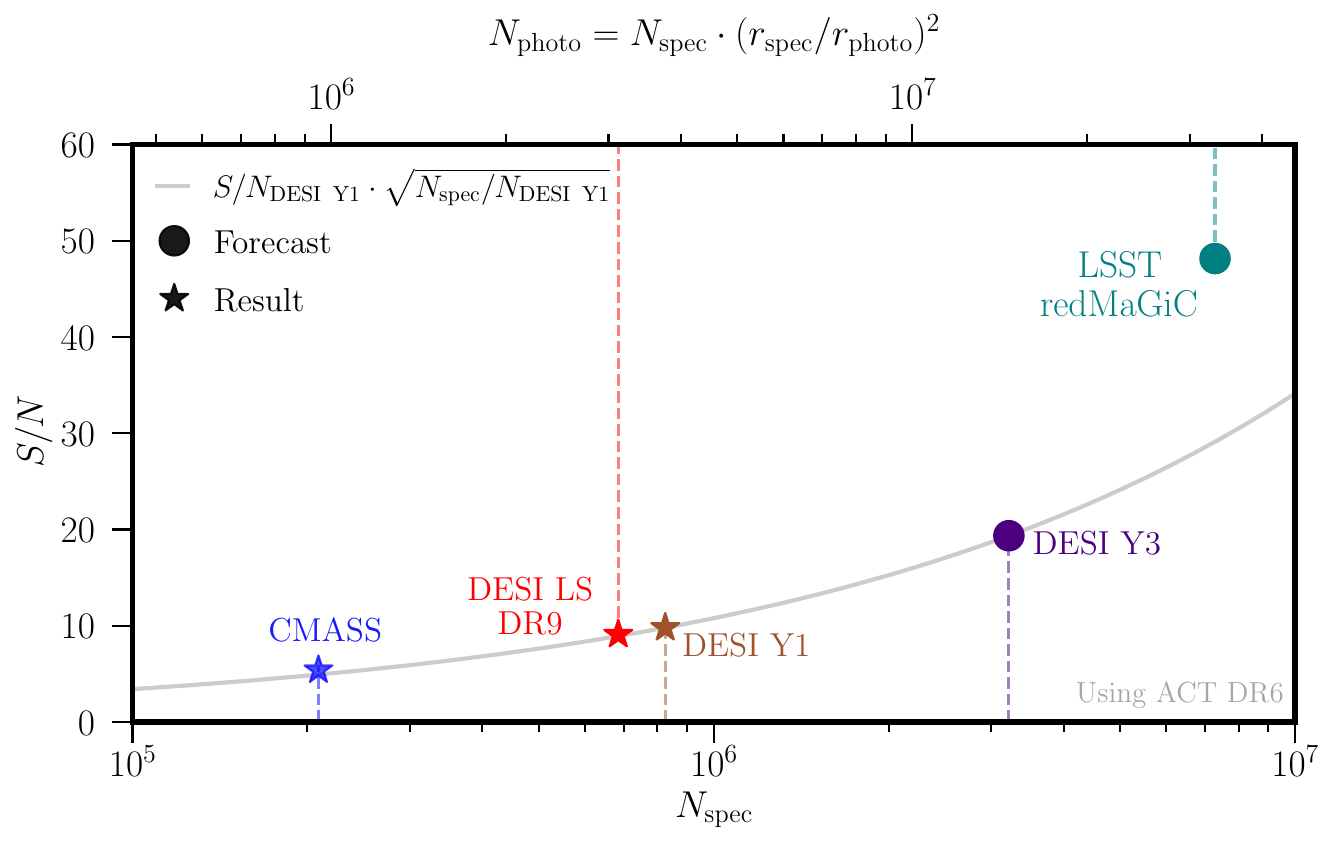}
    \caption{
    Results (stars symbols) and forecasts (circles) of the $S/N$ defined in Eq.~\ref{eq:SNR_null} for spectroscopic and photometric surveys with the ACT DR6 CMB map in the absence of clustering noise.
    As the number of galaxies increases, the $S/N$ grows proportional to the halo mass $M_{h}$, the correlation coefficient $r$ (defined in Eq.~\ref{eq:kSZ_stacking}), and the square root of the number of galaxies $N$.
    We included the curve $S/N_{\rm DESI \hspace{0.1 cm} Y1} \cdot \sqrt{N_{\rm spec}/N_{\rm DESI \hspace{0.1 cm} Y1}}$ to illustrate how the measured results from the CMASS and DESI samples are consistent with each other.
    For the forecasts, we find that the $S/N$ of the LSST redMaGiC selected galaxies and clusters will be slightly higher, which is due to a higher mean halo mass ($3.1 \cdot 10^{13} M_{\odot}/h$) as it includes clusters which we explicitly mask in this DESI LRG analysis to avoid tSZ contamination.
    We recall that the $S/N$ from DESI LS DR9 corresponds to a sub-sample of the one reported by \cite{hadzhiyska2024evidencelargebaryonicfeedback}.    
    }
    \label{fig:SNR_forecast}
\end{figure*}
%

In Fig. \ref{fig:SNR_forecast}, we present the $S/N$ of of present and future LRG catalogs.
To convert from an effective number of spectroscopic LRG to the equivalent photometric one in the context of measuring kSZ, we used the respective correlation coefficients and rescale $N_{\rm photo} = N_{\rm spec} \cdot (r_{\rm spec}/r_{\rm photo})^2$, assuming the mean masses are the same and $r_{\rm spec}/r_{\rm photo} = 0.65/0.30$ \cite{Hadzhiyska2023}.

The blue and brown star symbols correspond to the results shown in this work (CMASS and DESI Y1) when using the ACT DR6 CMB map.
The red star symbol (DESI LS DR9) shows the result when using the Legacy Survey Data Release 9 from the DESI collaboration \cite{hadzhiyska2024evidencelargebaryonicfeedback}.
To study how well this is consistent with 
\begin{equation}
    S/N \propto M_h \cdot r \cdot \sqrt{N},
    \label{eq:SNR_forecast}
\end{equation}
where $M_h$ corresponds to the host halo mass of the galaxy in units of $M_{\odot}/h$, $r$ to the correlation coefficient from Eq.~\ref{eq:kSZ_stacking}, and $N$ to the number of galaxies, we plot $S/N_{\rm DESI \hspace{0.1 cm} Y1} \cdot \sqrt{N_{\rm spec}/N_{\rm DESI \hspace{0.1 cm} Y1}}$ and find that the results are indeed well fitted by this relationship.

Given this consistency we extrapolate our results to forecast the $S/N$ of the next data release of LRG from DESI Year 3 (Y3)\footnote{LRG galaxies from DESI Year 5 will mainly consist of north-hemisphere samples, and therefore, would not overlap with the ACT map.}, and the selection of redMaGiC galaxies and clusters from the cosmoDC2 photometric simulation of LSST.
For the photometric forecasts, we assume the same mean photometric error ($\sigma(z)/(1+z) \lesssim 0.02$) \cite{Zhou_2023_2}, which matches the ``LSST goal'' \cite{lsstsciencecollaboration2009lsst}.

For the forecast of DESI Y3, in purple, we simply scaled by the number of LRG galaxies overlapping with ACT DR6 in the upcoming Y3 data release of DESI, corresponding approximately to $3.2 \cdot 10^6$ spectroscopically observed galaxies. 
This results in a kSZ $S/N^{\rm DESI \hspace{0.1 cm} Y3} \approx 20$, which will allow us to characterize with further detail the kSZ evolution for different parameters, among them the ones studied in the previous sections of this work.

For the LSST redMaGiC forecast, in teal, we used the extra-galactic catalog from the simulated LSST Data Challenge 2, named cosmoDC2 \cite{Korytov_2019, DC2_2021} and scaled them to account for the same footprint area that LSST will have\footnote{The simulation provides the sample from the Wide Fast Deep campaign for 5 years of observations (corresponding to Rubin’s DR6).}.
The LSST redMaGiC sample relies on the \texttt{redMaGiC} algorithm \cite{Rozo_2016}, based in \texttt{redMaPPer} \cite{Rykoff_2014}, which in principle selects LRG galaxies, but also clusters.
The mean mass of this sample is $3.1 \cdot 10^{13} M_{\odot}/h$, slightly larger than the mean halo mass of the LRG sample estimated using the AbacusHOD in the DESI One-Percent Survey \cite{Zhou2020}.
For 33 million redMaGiC objects, this would lead to an estimated $S/N$ $\sim 48$.
This result is above the curve due to the presence of clusters in the sample, which we otherwise mask for DESI to avoid tSZ contamination.

\section{Conclusion}
\label{sec:discussion_and_conclusion}

In this work, we measure the kSZ signal from the Y1 data release of the DESI LRG in combination with ACT DR6.
To do this, we stack the CMB map at the positions of the galaxies, 
weighting each galaxy by its peculiar LOS velocity estimated from the galaxy number density field.
We calculate the integrated kSZ measurement within compensated aperture photometry filters in order to reduce the CMB primary.
Our kSZ measurement with $S/N=9.8$ using spectroscopic LRGs is consistent with the photometric LRG measurement from  \cite{hadzhiyska2024evidencelargebaryonicfeedback}.
This increases our confidence in kSZ measurements from future photometric surveys, such as Rubin, Euclid and Roman.

We highlight a challenge in comparing data with hydrodynamical simulations.
Due to their finite volume, the simulations may not include the correct cosmological halo abundance at high masses, such that a simple abundance matching may not reproduce the correct halo mass distribution.
We explore two simple approaches to circumvent this issue, and leave a more thorough study to future work.
The first approach simply matches the amplitude of the simulated profiles at the largest measured aperture, where we expect all the baryons associated with the halo to be included.
This results in large discrepancies between both the dark matter and IllustrisTNG profiles and our data, as noted by \cite{hadzhiyska2024evidencelargebaryonicfeedback}.
However, this approach relies on the largest aperture measured, which is also the noisiest.
Our second approach frees the amplitude of the simulated gas profiles and fits it to the measurement.
As a result, a given model can only be ruled out this way based on a mismatch in profile shape.
This approach, which does not make use of the amplitude information in our measurement, is therefore conservative in this respect.
In this approach, the dark matter profile is again disfavored, though with a lesser confidence of 99.5\% (i.e. PTE = 0.005).
The IllustrisTNG model is only disfavored at 96.4\% confidence (i.e. PTE = 0.046).
Further investigation is warranted to substantiate these results.
Larger simulations, such as MillenniumTNG \cite{Pakmor_2023}, could be employed to forgo amplitude fitting, thereby providing a more robust basis for interpreting the observational results.

Additionally, we study the evolution of the kSZ profile with redshift, stellar mass, and absolute magnitude. 
We notice no clear trend with redshift,  but we find a distinct evolution when splitting the galaxies according to their stellar mass and apparent magnitude: the kSZ amplitude increases with larger gas mass.

We also measure the kSZ from BGS galaxies and ELG in the DESI Y1 data and found that their profiles correspond to their expected characteristics. 
These include an amplitude for the BGS similar to that found for the LRG due to a selection effect and a much more extended gas profile for the ELG, attributed to the large satellite fraction in that particular sample.
With these profiles, we prove it is possible to extend the kSZ study to other galaxy samples.

We present a forecast on the $S/N$ of future kSZ measurements of LRGs with ACT.
Our simple scaling with mass and number of objects matches existing measurements extremely well, and predicts a large $S/N$ with future datasets (e.g., DESI Y3, LSST with ACT DR6), significantly constraining uncertainties in the distribution of baryons and their evolution through different cosmic epochs.
Future CMB experiments, such as SO, CCAT and CMB-S4, will lead to further improvements.

We conclude that the $S/N$ in kSZ measurements is now large enough to allow us to conduct detailed analyses of the dependence of gas mass and feedback on galaxy properties. 
This promises to provide useful insights into feedback in galaxy formation and the calibration of sub-grid parameters in cosmological hydrodynamical simulations. 
The forthcoming further characterization of the kSZ measurements presented in this work and in \cite{hadzhiyska2024evidencelargebaryonicfeedback, Moser2022, https://doi.org/10.48550/arxiv.2307.10919, mccarthy2024flamingocombiningkineticsz, bigwood2024weak, bigwood2025caselargescaleagnfeedback}, represent exciting steps toward using kSZ to precisely localize baryons in halos of various masses and redshifts. 
This should further help calibrate baryonic effects in weak lensing, one of the biggest challenges currently facing observational cosmology.


\section*{Acknowledgments}

Data points for the figures are available at \url{https://doi.org/10.5281/zenodo.14908203}.

We thank ChangHoon Hahn and John Moustakas for guidance on the LRG mass estimates.
We also thank Theo Schutt for useful information on the DC2 catalog. 
We thank Sadaf Kadir for providing simulated gas profiles around DESI-like LRG in the IllustrisTNG simulation.
This work received support from the U.S. Department of Energy under contract number DE-AC02-76SF00515 to SLAC National Accelerator Laboratory.
This research used resources from the National Energy Research Scientific Computing Center (NERSC), a U.S. Department of Energy Office of Science User Facility located at Lawrence Berkeley National Laboratory.
B.H. is generously supported by the Miller Institute at University of California, Berkeley.
S.F. is supported by Lawrence Berkeley National Laboratory and the Director, Office of Science, Office of High Energy Physics of the U.S. Department of Energy under Contract No.\ DE-AC02-05CH11231.
C.S. acknowledges support from the Agencia Nacional de Investigaci\'on y Desarrollo (ANID) through Basal project FB210003.
D.N.S. is supported by the Simons Foundation.
K.M. acknowledges support from the National Research Foundation of South Africa.

Support for ACT was through the U.S.~National Science Foundation through awards AST-0408698, AST-0965625, and AST-1440226 for the ACT project, as well as awards PHY-0355328, PHY-0855887 and PHY-1214379. Funding was also provided by Princeton University, the University of Pennsylvania, and a Canada Foundation for Innovation (CFI) award to UBC. ACT operated in the Parque Astron\'omico Atacama in northern Chile under the auspices of the Agencia Nacional de Investigaci\'on y Desarrollo (ANID). The development of multichroic detectors and lenses was supported by NASA grants NNX13AE56G and NNX14AB58G. Detector research at NIST was supported by the NIST Innovations in Measurement Science program. Computing for ACT was performed using the Princeton Research Computing resources at Princeton University, the National Energy Research Scientific Computing Center (NERSC), and the Niagara supercomputer at the SciNet HPC Consortium. SciNet is funded by the CFI under the auspices of Compute Canada, the Government of Ontario, the Ontario Research Fund–Research Excellence, and the University of Toronto. We thank the Republic of Chile for hosting ACT in the northern Atacama, and the local indigenous Licanantay communities whom we follow in observing and learning from the night sky.

This material is based upon work supported by the U.S. Department of Energy (DOE), Office of Science, Office of High-Energy Physics, under Contract No. DE–AC02–05CH11231, and by the National Energy Research Scientific Computing Center, a DOE Office of Science User Facility under the same contract. Additional support for DESI was provided by the U.S. National Science Foundation (NSF), Division of Astronomical Sciences under Contract No. AST-0950945 to the NSF’s National Optical-Infrared Astronomy Research Laboratory; the Science and Technology Facilities Council of the United Kingdom; the Gordon and Betty Moore Foundation; the Heising-Simons Foundation; the French Alternative Energies and Atomic Energy Commission (CEA); the National Council of Humanities, Science and Technology of Mexico (CONAHCYT); the Ministry of Science, Innovation and Universities of Spain (MICIU/AEI/10.13039/501100011033), and by the DESI Member Institutions: \url{https://www.desi.lbl.gov/collaborating-institutions}. Any opinions, findings, and conclusions or recommendations expressed in this material are those of the author(s) and do not necessarily reflect the views of the U. S. National Science Foundation, the U. S. Department of Energy, or any of the listed funding agencies.

The authors are honored to be permitted to conduct scientific research on Iolkam Du’ag (Kitt Peak), a mountain with particular significance to the Tohono O’odham Nation.

\bibliographystyle{prsty.bst}
\bibliography{refs}


\appendix


\subsection{Skewness of the reconstructed velocity field}
\label{sec:skewness}

\begin{figure}
    \centering
    \includegraphics[width=0.48\textwidth]{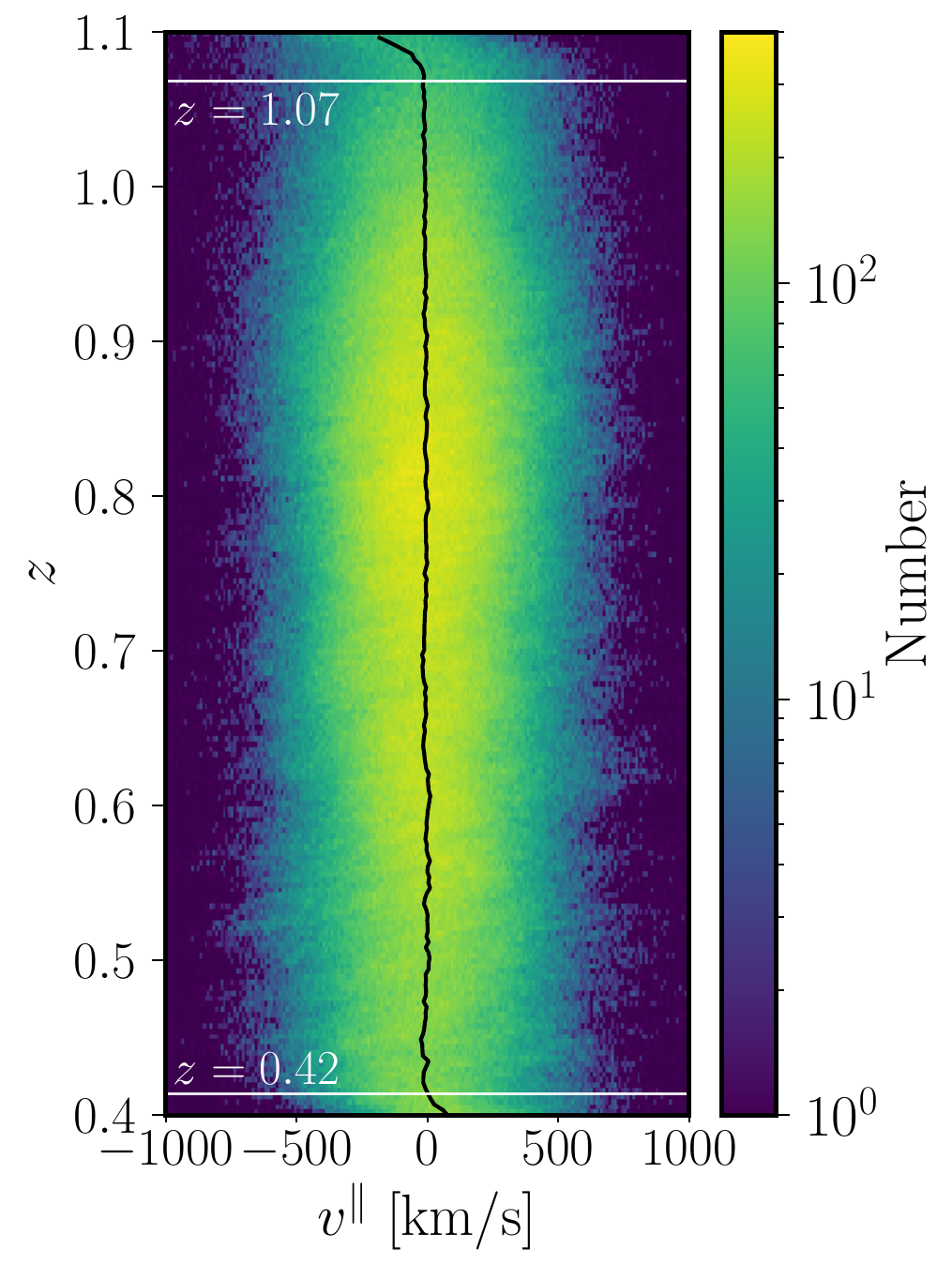}
    \caption{
    2D histogram in logarithmic scale of the reconstructed velocity along the LOS across redshift.
    In black, we plot the mean velocity per horizontal redshift bin, finding a positive and negative tendency for the lower and higher bounds respectively.
    The horizontal white lines indicate the boundaries where the mean peculiar velocity is zero.
        }
    \label{fig:skewness_velocity}
\end{figure}

Fig.~\ref{fig:vrec_vcent_vsat_1D_histogram} shows a slightly skewed distribution of LOS velocities (gray line), which we explore in this appendix.
When instead of plotting the 1D velocity distribution, we visualize the reconstructed velocities for different redshift bins as in Fig.~\ref{fig:skewness_velocity}, we find a survey volume effect.
Galaxies with redshifts less (more) than 0.42 (1.07), tend to have forward (recessive) velocities, as the mean (in black) tend to deviates from 0 km/s.
When not including these galaxies (19,737 and 20,797 galaxies respectively, which corresponds to $\sim$5$\%$ of the total sample), the kSZ $S/N$ changes by only $0.1$ and the velocity distribution is no longer skewed (black line from Fig.~\ref{fig:vrec_vcent_vsat_1D_histogram}).
This is expected as the number of galaxies at those extreme redshifts is low compared to the whole dataset.


%
\begin{figure}
    \centering
    \includegraphics[width=0.48\textwidth]{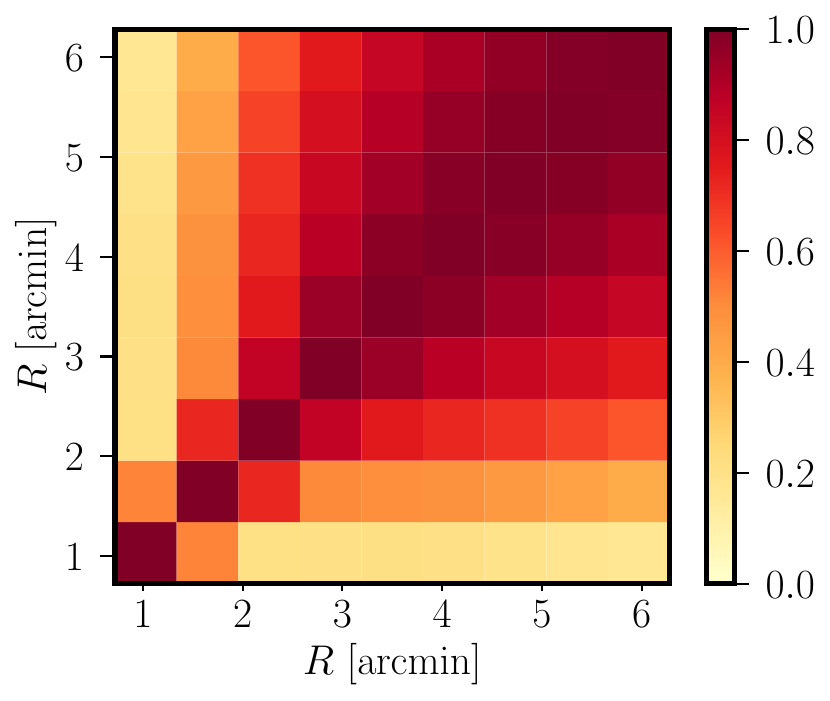}
    \caption{
     Correlation coefficient covariance matrix between the different CAP filter radius for the kSZ fiducial measurement.
     The correlation is stronger on large scales, as the fluctuations of the primary CMB become dominant, characteristic in all kSZ measurements of this work.}
    \label{fig:cor_diskring_ksz_uniformweight_bootstrap}
\end{figure}

\subsection{Covariance matrix}
\label{sec:cov_matrix}

To estimate the covariance matrices, we bootstrap-resample individual galaxies (with repetition) until the number of galaxies matches that in the catalog. 
This process was then repeated 10,000 times, generating resampled and scattered galaxy catalogs. 
The covariance matrices were computed from the stacked profiles of these resampled catalogs.
In the limit of independent noise realization of the galaxies, which is close to accurate, this assumption works, as shown by \cite{Schaan2021}.
Eventually the larger angular scales of the CMB cutouts do overlap, and therefore, we must take that effect into account.

The covariance matrix of the fiducial kSZ measurement of this work is shown in Fig.~\ref{fig:cor_diskring_ksz_uniformweight_bootstrap}.
The small scales are uncorrelated as the uncertainty is mainly coming from the CMB map noise.
However, larger scales start to be correlated as the CMB primary anisotropies start to be relevant.
Due to this, and following \cite{Schaan2021, hadzhiyska2024evidencelargebaryonicfeedback}, we set the maximum aperture to 6 arcmin.
All of our covariance matrices look similar to the one shown here.

\subsection{Best-fit ratio between two kSZ profiles}
\label{sec:matched_filter_aperture}

\begin{figure}
     \centering
     \begin{subfigure}[b]{0.48\textwidth}
         \centering
         \includegraphics[width=0.99\textwidth]{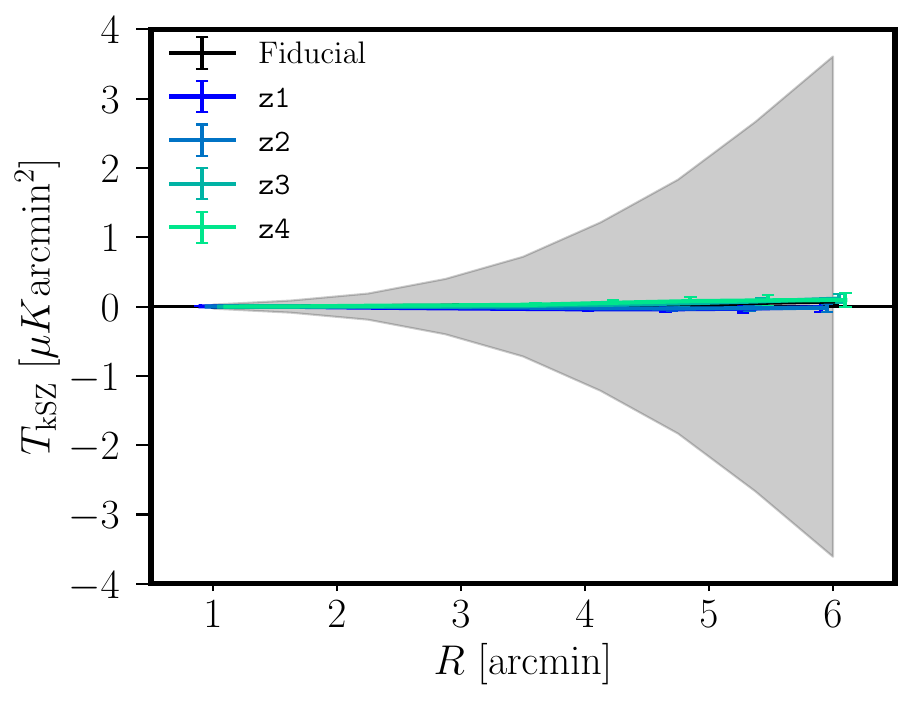}
     \end{subfigure}
     \hfill
     \begin{subfigure}[b]{0.48\textwidth}
         \centering         \includegraphics[width=0.99\textwidth]{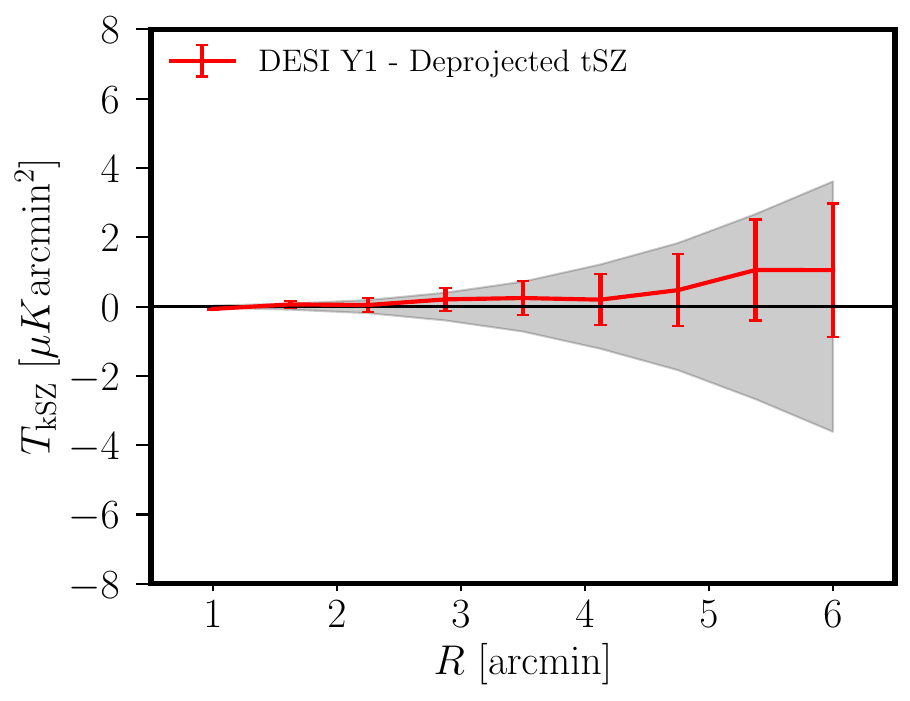}
     \end{subfigure}
     \caption{
     Systematic checks/Null tests:
     \textit{Upper panel:}
     Velocity reconstruction null test: 
     We check there are no residual CIB or tSZ contamination by randomly shuffling the velocities of the galaxies before measuring the kSZ. 
     In all the cases (the fiducial and when splitting into redshift bins), the results were consistent with 0.
     The gray band corresponds to the statistical error of the measurement.
     \textit{Bottom panel:}
     Thermal Sunyaev-Zel'dovich test: contamination null test: 
     Instead of stacking the hILC CMB map at the positions of the DESI LRG, we stack on the hILC CMB with deprojected tSZ map, which should not result in a substantial difference.
     Here we plot the difference between those two stacks proving that the tSZ is not filtering through and therefore causing a bias.  
     As in the upper panel, the gray band corresponds to the statistical error of the fiducial measurement.
     }
     \label{fig:syst_checks}
\end{figure}

Consider a data profile, denoted by the vector $\vec{d}$, and the fiducial curve from Fig.~\ref{fig:kSZ_signal} (represented by the vector $\vec{m}$) as our reference model. 
We then define an amplitude $A$ such that these quantities are related by:
\begin{equation}
    \vec{d} = A \cdot \vec{m} + \vec{n},
\end{equation}
where $\vec{n}$ the noise.
The corresponding likelihood with independent noise is
\begin{equation}
    \ln \mathcal{L}(A) = -\frac{1}{2}[\vec{d} - A \cdot \vec{m}]^{\top} {\rm C}^{-1}[\vec{d} - A \cdot \vec{m}].
\end{equation}
The amplitude best-fit is
\begin{equation}
    \hat{A} = \frac{\vec{m}^{\top} {\rm C}^{-1} \vec{d}}{\vec{m}^{\top} {\rm C}^{-1} \vec{m}},
    \label{eq:amplitude_matched_filter}
\end{equation}
and its noise estimate is
\begin{equation}
    \sigma_{A} = [\vec{m}(\vec{\theta})^{\top} {\rm C}^{-1} \vec{m}(\vec{\theta})]^{-1/2}.
    \label{eq:error_matched_filter}
\end{equation}
%


\subsection{Systematic checks/Null tests}
\label{sec:syst_checks}

\begin{figure}
    \centering
    \includegraphics[width=0.45\textwidth]{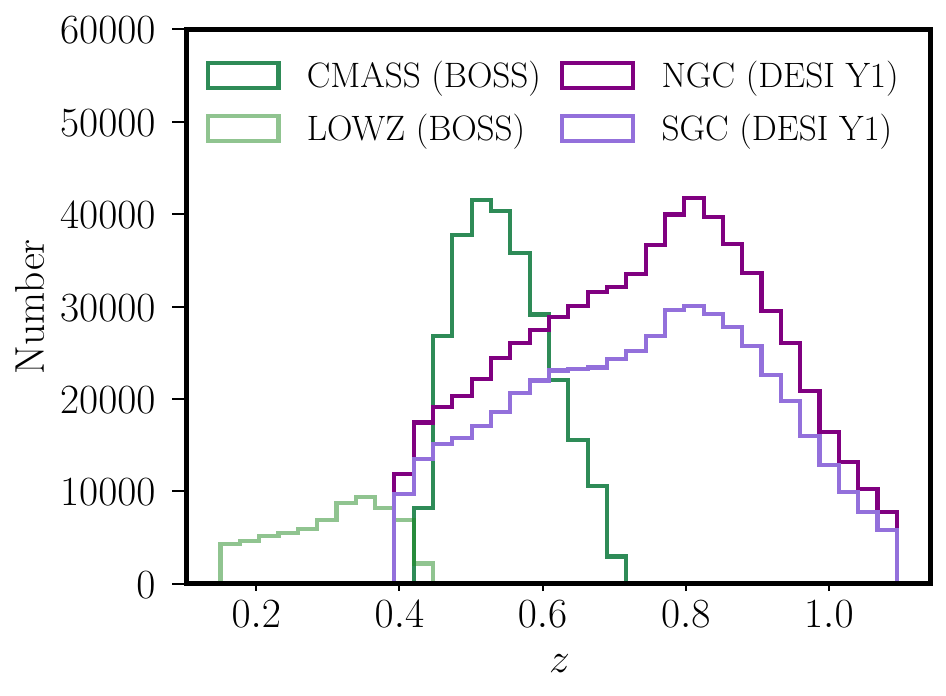}
    \caption{
    The redshift distribution of the DESI LRG Y1 is shown for both the North Galactic Cap (NGC) and South Galactic Cap (SGC), along with the BOSS data. 
    The NGC and SGC samples contain 1,476,132 and 662,468 galaxies, respectively. 
    For comparison, the CMASS and LOWZ samples from BOSS consist of 777,202 and 218,905 galaxies, respectively, and correspond to lower redshifts. 
    We emphasize that these are not the galaxies overlapping with the ACT DR6 map.}
    \label{fig:z_dist}
\end{figure}
\begin{figure}
    \centering
    \includegraphics[width=0.48\textwidth]{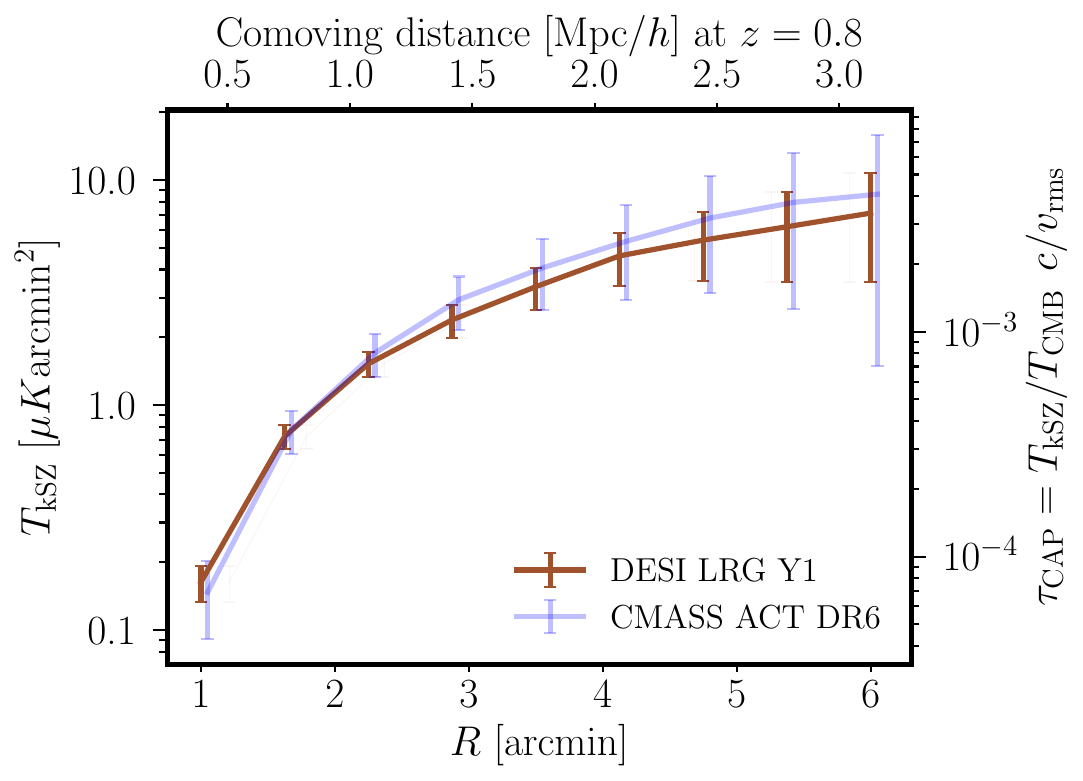}
    \caption{
    Similar to Fig.~\ref{fig:kSZ_signal}, the kSZ stacked profiles of both the DESI Y1 and CMASS LRG on the ACT DR6 map, for which we find great agreement.
    }
    \label{fig:DESI_vs_CMASS}
\end{figure}

To validate our measurement and show that correlated contaminants, such as the cosmic infrared background or tSZ, are not impacting our result, we do the two following systematic checks.

First, we randomly shuffle the reconstructed velocities of our galaxies and obtain 1000 realizations.
This would guarantee that we are indeed obtaining a measurement from kSZ that includes enough galaxies, such that the positive and negative weighting from the velocities cancels out other secondary anisotropies.
We stacked on these shuffled velocities and find no signal, as shown in the upper panel of Fig.~\ref{fig:syst_checks}.
This means that the tSZ and CIB effects did not bias our measurement and that the CMB map used is accurately separated.

Second, we stacked on the hILC deprojected CMB map, to check for possible tSZ or dust contamination.
In the lower panel of Fig.~\ref{fig:syst_checks} we show the difference between the fiducial kSZ measured from this work and the deprojected one.
We find no contamination, as the red curve is inside the statistical errors of the fiducial study.


\subsection{Agreement with CMASS x ACT results}
\label{sec:comparison_cmass}

\begin{table}
	\centering
	\begin{tabular}{|c|c|c|}
		\hline
            Parameter & CMASS M & DESI \\
            \hline
            $z_{\rm min}$ - $z_{\rm max}$ & 0.4 - 0.7 & 0.4 - 1.1 \\
            $z_{\rm eff}$ & 0.57 &  0.78 \\
            $N_{\rm gal}^{\rm total}$ & 777,202 & 2,138,600 \\
            $N_{\rm gal}^{\rm overlap}$ & 208,921 & 825,283 \\
            $r_s$ [Mpc/$h$] & 15 & 15 \\
            \hline
            $h$ &  0.676 & 0.6736 \\
            $\Omega_m$ & 0.31 & 0.31519 \\
            $b$ & 2.1 & 2.0 \\
            \hline
            $\sigma_{\rm rec}$ [km/s] & 204 & 233 \\
            \hline
	\end{tabular}
     \caption{
    Comparison between the velocity reconstruction on DESI and the CMASS M galaxies.
    The fiducial cosmology from the CMASS M sample is listed at \cite{Vargas_Maga_a_2018, Alam_2017}, while the DESI sample follows the Planck 2018 base-$\Lambda$CDM parameters \cite{Planck2020, 2024Paillas}.
    In \cite{Ried_Guachalla_2024}, we showed the difference between these cosmological parameters produces no substantial impact on the velocity reconstruction. 
    }
\label{tab:fiducial_parameters}
\end{table}

The kSZ stacking measurement done by \cite{Schaan2021} used CMASS galaxies and the ACT DR5 map.
In this section, we compare the kSZ measurements when using either CMASS or DESI LRG Y1 and the ACT DR6 map.
One first difference is the reduction of eligible CMASS galaxies to stack on, as the latest ACT map includes a more conservative mask that removes not previously considered point sources and clusters.
In total, 208,969 CMASS LRG galaxies overlap with the ACT DR6 map, which is equivalent to 42$\%$ less galaxies than what was used in \cite{Schaan2021} (see Fig.~\ref{fig:z_dist} for its redshift distribution).
We show the resulting profile in faint blue in Fig.~\ref{fig:DESI_vs_CMASS}, which is in agreement with DESI Y1.
The kSZ measurement results in $S/N^{\rm CMASS} = 5.4$.
The agreement with the DESI Y1 result is expected as the velocity reconstruction parameters and mean masses are similar, as shown in Table~\ref{tab:fiducial_parameters}.
The main difference is their redshift range and the final standard deviation of their reconstructed velocities.

In Fig.~\ref{fig:SNR_forecast} we include the $S/N^{\rm CMASS}$ with a blue star, finding it is in agreement with what would we expect if we sub-sample DESI to match the number of CMASS galaxies (gray line).

\end{document}